\documentclass[prd,final,aps]{revtex4}

\usepackage{graphicx}
\usepackage{amsmath,amsfonts,amstext,amssymb,amsbsy,amsopn}

\usepackage{yfonts}
\usepackage{accents}

\newcommand{\g}{\textgoth{g}}

\newcommand{\h}[1][\relax]{\ifx#1\relax \bar{h}  \else  \underaccent{\mathsf{#1}}{\h}  \fi}

\newcommand{\tPsi}[1][\relax]{\ifx#1\relax  {\widetilde{\Psi}}  \else  \underaccent{\,\mathsf{#1}}{\tPsi}  \fi}

\usepackage{amssymb}

\makeatletter

\newcommand{\mathbox}[2][normal]{\begingroup\ensuremath
   \let\@nomath\@gobble  \mathversion{#1}
   \mathchoice
     {\hbox{$\m@th\displaystyle      #2$}}
       {\hbox{$\m@th\textstyle         #2$}}
       {\hbox{$\m@th\scriptstyle       #2$}}
       {\hbox{$\m@th\scriptscriptstyle #2$}}
   \endgroup}

\DeclareFontShape{OT1}{cmss}{bx}{sl}{<-> cmssbxo10}{}

\DeclareMathVersion{boldoblique}
\SetSymbolFont{operators}{boldoblique}{OT1}{cmr}{bx}{sl}
\SetSymbolFont{letters}{boldoblique}{OML}{cmm}{b}{it}
\SetSymbolFont{symbols}{boldoblique}{OMS}{cmsy}{b}{n}
\SetMathAlphabet\mathsf{boldoblique}{OT1}{cmss}{bx}{sl}
\SetMathAlphabet\mathit{boldoblique}{OT1}{cmr}{bx}{it}

\makeatother

\usepackage{subfigure}

\newcommand{\bbla}{\;\;\;\;}

\advance\textwidth -.5in

\begin{document}

\bibliographystyle{prsty} 

\title{The periodic standing-wave approximation: 
computations in full general relativity}

\author{Napole\'on Hern\'andez} \affiliation{Department of Physics,
  University of Utah, Salt Lake City, UT 84112, and Department of
  Physics \& Astronomy and Center for Gravitational Wave Astronomy,
  University of Texas at Brownsville, Brownsville, TX 78520}
\altaffiliation{current address: JP Morgan, 277 Park Avenue, Floor
12, New York, NY 10172-0003.
}
\author{Richard H.~Price} 
\affiliation{Department of Physics \& Astronomy and Center for 
Gravitational Wave Astronomy, University of Texas at Brownsville,
Brownsville, TX 78520}

\begin{abstract}
\begin{center}
{\bf Abstract}
\end{center}
The periodic standing wave method studies circular orbits of compact
objects coupled to helically symmetric standing wave gravitational
fields. From this solution an approximation is extracted for the
strong field, slowly inspiralling motion of binary black holes and
binary neutron stars. Previous work on this project has developed a
method using a few multipoles of specially adapted coordinates well
suited both to the radiation and the source regions. This method had previously
been applied to linear and nonlinear scalar field models, to
linearized gravity, and to a post-Minkowski approximation. Here we
present the culmination of this approach: the application of the
method in full general relativity. The fundamental equations had
previously been developed and the challenge presented by this step is
primarily a computational one which was approached with an innovative
technique. The numerical results of these computations 
are compared with the
corresponding results from linearized and post-Minkowksi computations.

\end{abstract}
\maketitle

\section{Background and introduction}\label{sec:intro} 

Due to successes in numerical relativity, there is now a good
understanding of many features of the gravitational waves from the
inspiral and merger of comparable mass black holes\cite{pretorius1,
pretorius2,
pretoriuskhurana,
pretorius3,
utbpuncturePRL,
utbpuncturePRD,
rochestermax,
goddardpuncturePRL,
goddardpuncturePRD,
goddardkicks,
goddardcharacteristics,
goddardmodelkicks, 
compare,
dienerPRL,
jenakicks,
dienerkicks,
jenafinalspin, 
jenaeccentric}
There are, however, areas that are still best investigated with an
approximation method for the slow inspiral epoch.  One area is the
evolution and gravitational waveform during the slow inspiral (but
strong field) epoch, since there are too many orbits for numerical
relativity to be feasible.  Another area is the case of mass ratios
too large for numerical relativity, but not large enough for particle
perturbation methods. A slow inspiral approximation method is also
useful as a starting point for a better understanding of the radiation
reaction fields acting on an inspiralling hole.

The Periodic Standing Wave (PSW) project is 
meant to provide such an approximation.  This method seeks a numerical
solution for a pair of sources (black holes, neutron stars) in
nondecaying circular orbits with  gravitational fields that are
rigidly rotating, that is, fields that are helically
symmetric. Because the universality of gravitation will not permit
outgoing waves and nondecaying orbits, the solution to be computed is
that for standing waves. An approximation for slowly decaying orbits
with outgoing radiation is then extracted from that numerical
solution.

This work has progressed through several stages. In the first
stage\cite{WKP,WBLandP, rightapprox,paperI,eigenspec}, a nonlinear
scalar fields model was investigated, and numerical methods were
developed to deal with the special mathematical features that would be
common to all standing-wave, helically symmetric computations.  These
features include: (i)~a mixed boundary value problem (regions of the
domain in which the equations are hyperbolic and other regions in
which they are elliptic); (ii)~an iterative construction of nonlinear
standing wave solutions; (iii)~the extraction from the standing wave
solution of an approximate outgoing wave solution. 
Reference \cite{eigenspec} introduced a new technique that promised to
reduce the computational burden: ``adapted coordinates'' that were
well suited to the geometry of the problem both near the sources and
far from them. In Ref.~\cite{eigenspec} it was shown that with these
coordinates good results could be computed by keeping only a very
small number of multipoles, typically just the monopole and quadrupole
moments. That method was applied to linearized general relativity in
Ref.~\cite{lineareigen} and to the post-Minkowski approximation to
general relativity in Ref.~\cite{pmpaper}. In the current paper we
apply this method, PSW in adapted coordinates, to full general
relativity.

The mathematical infrastructure for this is rather involved. It
includes the definition of standing waves for nonlinear fields;
extraction of the outgoing solution; the adapted spatial coordinates;
the multipole decomposition in adapted coordinates; the reduction to
helical symmetry for tensorial fields; the full vacuum field
equations; the inner boundary conditions on surfaces outside black
holes. This infrastructure has been thoroughly documented in earlier
papers, and the complete infrastructure is presented in
Ref.~\cite{pmpaper}. Rather than give yet another lengthy presentation
of this infrastructure, here we will give only
enough of the mathematical background necessary in order to explain
the numerical results we are presenting, and to compare them with
other numerical results.  

The rest of the paper is organized as
follows. Sec.~\ref{sec:formalism} develops the mathematical
description of helically symmetric tensor fields in terms of ``helical
scalars,'' quantities that are functions only of corotating
coordinates. The field equations of general relativity, and of its
weak field approximations are given for these helical scalars, along
with boundary conditions on these fields. Section \ref{sec:nummeth}
discusses two important aspects of the numerical approach to our
computations: (i)~``discrete spherical harmonics'' on our
computational grid, and (ii)~the use of the \emph{Maple}\texttrademark
symbolic manipulation language to handle the complexity of the
equations to be solved numerically.  The numerical results produced by
these methods are presented in Sec.~\ref{sec:numresults}, and are
discussed in Sec.~\ref{sec:conc}.

\section{The mathematical problem}\label{sec:formalism}

\subsection{The field equations and coordinates}\label{sec:fieldequations}

We start with the concept of a strong field source region and a weak
field source region. The strong field source region is close to the
inner computational boundary, surfaces with spherical topology around
the black holes, and close to them but far enough from the horizon for
computations to be feasible.  We invoke a set of coordinates labelled
$t,x,y,z$ for which the spacetime metric $g_{\mu\nu}$ deviates only
slightly from the flat metric $\eta_{\mu\nu}$ at large coordinate
distances from the binary sources.

We impose helical symmetry by requiring that in these coordinates
there be a Killing vector given by
\begin{equation}\label{eq:xidef0} 
\xi=\partial_t+\Omega\left(x\partial_y-y\partial_x
\right)
\end{equation}
where $\Omega$
is a constant.
It is convenient to define a rotating coordinate
system by
\begin{equation}
\widetilde{t}={t}\quad\quad
\widetilde{z}={z}\quad\quad
\widetilde{x}\equiv x\,\cos{\Omega t}+y\sin{\Omega t}
\quad\quad
\widetilde{y}\equiv \,-x\sin{\Omega t}+y\cos{\Omega t}\,.
\end{equation} 
and  to introduce two cylindrical 
coordinates sytems: $r,z,\phi$ in terms of $x,y,z$ by the usual flat space
formulas, and 
$\widetilde{r},\widetilde{z},\widetilde{\phi}$ in terms of $\widetilde{x},
\widetilde{y},\widetilde{z}$ also by the usual flat space
formulas.
We note that Eq.~(\ref{eq:xidef0}) is equivalent to
\begin{equation}\label{eq:xidefalt} 
\xi=\partial_t+\Omega\left(\widetilde{x}\partial_{\widetilde{y}}
-\widetilde{y}\partial_{\widetilde{x}}
\right)=\partial_t+\Omega\partial_\varphi
=
\partial_{\widetilde{t}}\,,
\end{equation}

We follow the formulation of Landau and Lifschitz 
\cite{landaulifschitz,MTWsec20.3} for the Einstein equations, which
encodes the geometric information in the densitized 
metric
\begin{equation}
\g^{\alpha\beta}\equiv \sqrt{|{\det g}|}\, g^{\alpha\beta}\,.\label{eq:gDef}
\end{equation}
To simplify the field equations, we
choose to have our coordinates obey the harmonic condition
\begin{equation}\label{eq:hgGauge}
\g^{\alpha\beta}{}_{,\beta} = 0\,,
\end{equation}
we define the inverse $\g_{\alpha\beta}$ of our basic field
$\g^{\alpha\beta}$ by
\begin{equation}\label{eq:gInv}
\g^{\alpha\beta}\, \g_{\beta\gamma} = \delta^\alpha{}_\gamma \,,
\end{equation}
and we define $\h^{ab} $ by
\begin{equation}\label{eq:hdef} 
\g^{\alpha\beta}
\equiv \sqrt{-\det \eta}\, \bigl(\eta^{\alpha\beta}-\h^{\alpha\beta} \bigr).
\end{equation}
The vacuum field equations then take the simple form
\begin{equation}\label{eq:llheq}
 \Box\h^{\alpha\beta}\equiv \eta^{\rho\sigma}\,
\h^{\alpha\beta}{}_{,\rho\sigma} = S^{\alpha \beta \kappa \nu}_{\tau \phi \lambda \mu}\; 
\h^{\tau\phi}{}_{,\kappa}\, \h^{\lambda\mu}{}_{,\nu}+\h^{\rho\sigma}\h^{\alpha\beta}{}_{,\rho\sigma}\,,
\end{equation}
where
\begin{multline}\label{fullS}
S^{\alpha \beta \kappa \nu}_{\tau \phi \lambda \mu}\;=-\Bigl[ \delta^{(\alpha}_\rho\,
\delta^{\beta)}_\sigma - \textstyle{\frac{1}{2}}\, \g^{\alpha\beta}\,
\g_{\rho\sigma} \Bigr] \Bigl[ \delta^\rho_\tau\, \delta^\sigma_\lambda\,
\delta^\nu_\phi\, \delta^\kappa_\mu - 2\, \delta^\rho_\tau\, \g^{\sigma\nu}\, \g_{\phi\lambda}\,
\delta^\kappa_\mu + \delta^\rho_\tau\, \delta^\sigma_\lambda\, \g_{\phi\mu}\, \g^{\kappa\nu} \\
		- \textstyle{\frac{1}{2}}\, \g^{\rho\sigma}\, \g_{\tau\lambda}\,
\g_{\phi\mu}\, \g^{\kappa\nu} + \textstyle{\frac{1}{4}}\, \g^{\rho\kappa}\, \g^{\sigma\nu}
\Bigl( 2\g_{\tau\lambda}\, \g_{\phi\mu} - \g_{\tau\phi}\, \g_{\lambda \mu} \Bigr) \Bigr]\,.
\end{multline}
Equation (\ref{eq:llheq}), along with the definition 
(\ref{eq:hdef}), are the  equations to be solved for the unknown 
fields $\h^{\alpha\beta}$. 

Note that so far Eq.~(\ref{eq:llheq}) 
is exact; there is no assumption that the 
$\h^{\alpha\beta}$ fields are perturbative. If we do treat
these fields as perturbative,  the post-Minkowski approximation follows
by replacing $\g$ on the right by $\eta$ to get
\begin{multline}\label{eq:generalpm}
 \Box\h^{\alpha\beta}\equiv \eta^{\rho\sigma}\,
\h^{\alpha\beta}{}_{,\rho\sigma} = -\Bigl[ \delta^{(\alpha}_\rho\,
\delta^{\beta)}_\sigma - \textstyle{\frac{1}{2}}\, \eta^{\alpha\beta}\,
\eta_{\rho\sigma} \Bigr] \Bigl[ \delta^\rho_\tau\, \delta^\sigma_\lambda\,
\delta^\nu_\phi\, \delta^\kappa_\mu - 2\, \delta^\rho_\tau\, \eta^{\sigma\nu}\, \eta_{\phi\lambda}\,
\delta^\kappa_\mu + \delta^\rho_\tau\, \delta^\sigma_\lambda\, \eta_{\phi\mu}\, \eta^{\kappa\nu} \\
		- \textstyle{\frac{1}{2}}\, \eta^{\rho\sigma}\, \eta_{\tau\lambda}\,
\eta_{\phi\mu}\, \eta^{\kappa\nu} + \textstyle{\frac{1}{4}}\, \eta^{\rho\kappa}\, \eta^{\sigma\nu}
\Bigl( 2\eta_{\tau\lambda}\, \eta_{\phi\mu} - \eta_{\tau\phi}\, \eta_{\lambda \mu} \Bigr) \Bigr]
\h^{\tau\phi}{}_{,\kappa}\, \h^{\lambda\mu}{}_{,\nu}+\h^{\rho\sigma}\h^{\alpha\beta}{}_{,\rho\sigma}\,.
\end{multline}
For linearized general relativity, the terms on the right can be ignord and the
field equations are simply
\begin{equation}
  \label{eq:linorder}
\Box\h^{\alpha\beta}=0\,.
\end{equation}
These can be solved in closed form as expansions in special
functions\cite{lineareigen}.

In the case of a scalar field model, helical symmetry means that the
scalar field $\Phi$ is a function only of corotating coordinates
$\widetilde{x},\widetilde{y},\widetilde{z}$, or
$\widetilde{r},\widetilde{\phi},\widetilde{z}$, and not of
$\widetilde{t}$. The computations for the field then involve only
three coordinates. Complications arise when dealing with the
components of tensor fields, since computational fields must be
``helical scalars,'' i.e.\,, functions only of the rotating
coordinates. The approach we employ is to define
\begin{eqnarray}
\widetilde{\Psi}^{(nn)}&=&\bar{h}^{tt}\label{eq:Psifirst}\\
\widetilde{\Psi}^{(n0)}&=&\textstyle{\sqrt{2\;}}\ \bar{h}^{tz}\\
\widetilde{\Psi}^{(n1)}&=&e^{i\Omega t}\left(\bar{h}^{tx}-i\bar{h}^{ty}\right)\equiv U^{(n1)}+iV^{(n1)}
=-\left(\widetilde{\Psi}^{(n,-1)}\right)^*
\\
\Psi^{(00)}&=&\textstyle{\frac{1}{\sqrt{3\;}}}\left[
\bar{h}^{xx}+\bar{h}^{yy}+\bar{h}^{zz}
\right]\\
\widetilde{\Psi}^{(20)}&=&\textstyle{\frac{-1}{\sqrt{6\;}}}\left[
\bar{h}^{xx}+\bar{h}^{yy}-2\bar{h}^{zz}
\right]\\
\widetilde{\Psi}^{(21)}&=&e^{i\Omega t}
\left(-
\bar{h}^{xz}
+i\bar{h}^{yz}\right)\equiv U^{(21)}+iV^{(21)}=-\left(\widetilde{\Psi}^{(2,-1)}\right)^*\\
\widetilde{\Psi}^{(22)}&=&e^{2i\Omega t}\left(\textstyle{\frac{1}{2}}\left[
\bar{h}^{xx}-\bar{h}^{yy}\right]
-i\bar{h}^{yx}\right)\equiv U^{(22)}+iV^{(22)}
=\left(\widetilde{\Psi}^{(2,-2)}\right)^*\,,
\label{eq:Psilast}
\end{eqnarray}
in which $\widetilde\Psi^{(nn)}$,
$\widetilde\Psi^{(n0)}$,
$U^{(n1)},V^{(n1)}$,
$\widetilde\Psi^{(00)}$,
$\widetilde\Psi^{(20)}$,
$U^{(21)},V^{(21)}$,
$U^{(22)},V^{(22)}$,
are 10 real, helical scalars that carry all the information about the metric.

We denote any of  4 real and 3 complex helical scalars as
$\widetilde{\Psi}^{A}$, with $A$ taking the value $nn, n0, n1,...$.
In terms of these helical scalars the field equations (\ref{eq:llheq})
take the form
\begin{equation}\label{boxComplex} 
\Box\widetilde{\Psi}^{A}
-2i\mu(A)\Omega^2\partial_{\varphi}\widetilde{\Psi}^{A}
+\mu(A)^2
\Omega^2\widetilde{\Psi}^{A}=
{\cal Q}^{A}(\widetilde{\Psi}^B)\,,
\end{equation}
where $\mu(A)$ has the value of $0$ for $A=(nn),(n0),(00),(20)$, has
the value $\pm 1$ for $A=(n\pm 1), (2\pm 1)$ and has value $\pm 2$ for
$A=(2\pm 2)$. 

In order to give the details of the source term we must introduce a
set of  objects ${\bf n}, {\bf
  e}_x,{\bf e}_y,{\bf e}_z $. These objects are described in detail in Ref.~\cite{pmpaper},
and can roughly be considered to be the  basis vectors associated with coordinates $t,x,y,z$. 
In terms of these, we define the objects
\begin{eqnarray}
\widetilde{{\bf t}}_{nn}&\equiv&{\bf n}{\bf n}\label{tnndef}  \\
\widetilde{{\bf t}}_{n0}&\equiv& \textstyle{\frac{1}{\sqrt{2\;}}}\left[{\bf n}{\bf e}_z
+{\bf e}_z{\bf n}
\right]\\
\widetilde{{\bf t}}_{n,\pm1}&\equiv&
e^{\mp i\Omega t}
\left(\textstyle{\frac{\mp1}{{2\;}}}\right)\left[
{\bf n}({\bf e}_x\pm i{\bf e}_y)
+({\bf e}_x\pm i{\bf e}_y){\bf n}
\right]
\\
\widetilde{{\bf t}}_{0,0}&\equiv&\textstyle{\frac{1}{\sqrt{3\;}}}
\left[{\bf e}_x{\bf e}_x+{\bf e}_y{\bf e}_y+{\bf e}_z{\bf e}_z\right]\\
\widetilde{{\bf t}}_{2,0}&\equiv&\textstyle{\frac{-1}{\sqrt{6\;}}}
\left[{\bf e}_x{\bf e}_x+{\bf e}_y{\bf e}_y-2{\bf e}_z{\bf e}_z\right]\\
\widetilde{{\bf t}}_{2,\pm1}&\equiv&e^{\mp i\Omega t}\left(\mp\textstyle{\frac{1}{2}}\right)
\left[{\bf e}_x{\bf e}_z+{\bf e}_z{\bf e}_x\right]
-\textstyle{\frac{1}{2}}\,i
\left[{\bf e}_y{\bf e}_z+{\bf e}_z{\bf e}_y\right]\\
\widetilde{{\bf t}}_{2,\pm2}&\equiv&e^{\mp 2i\Omega t}\textstyle{\frac{1}{2}}
\left[{\bf e}_x{\bf e}_x-{\bf e}_y{\bf e}_y\pm i\left({\bf e}_y{\bf e}_x+{\bf e}_x{\bf e}_y
\right)
\right]\,,xs\label{t22def}
\end{eqnarray}
which are shown to be helical scalars in Ref.~\cite{pmpaper}.
With this notation the explicit form of the source is
 \cite{napthesis}
\begin{multline}
{\cal Q}^{A}(\widetilde{\Psi}) =
  \left(\tilde{t}^{\alpha \beta}_A\right)^* 
S^{\alpha \beta
    \kappa \nu}_{\tau \phi \lambda \mu}\; \bar{t}^{\tau \phi}_B
  \bar{t}^{\lambda \mu}_C \left[ \tilde{\Psi}^B_{,\kappa} -
    i\mu(B)\delta^t_{\kappa} \tilde{\Psi}^B \right] \; \left[
    \tilde{\Psi}^C_{,\nu} - i\mu(C)\delta^t_{\nu} \tilde{\Psi}^C
  \right] \\ +\bar{t}^{\rho \sigma}_B \tilde{\Psi}^B
  \left[\tilde{\Psi}^A_{,\rho \sigma} - i \mu(A) \Omega
    \left(\delta^{t}_{\rho}
      \tilde{\Psi}^A_{,\sigma}+\delta^{t}_{\sigma}
      \tilde{\Psi}^A_{,\rho}\right)-\mu(A)^2\Omega^2 \delta^{t}_{\rho}
    \delta^{t}_{\sigma}\tilde{\Psi}^A\right]. \label{Qterm}
\end{multline}
Here the summation is not only over the tensorial indices, but also
over the indices,$B$, $C$ which range over the 10 values
$(nn),(n0),(n\pm1),(00),(20),(2\pm1),(2\pm2) $.

Rather than work with the complex fields $\widetilde{\Psi}^{(n1)},
\widetilde{\Psi}^{(21)},\widetilde{\Psi}^{(22)}$,
in practice we work with the real and imaginary 
parts $U^{A}$ and $V^{A}$, and for $A=(n1),(21),(22)$ 
Eq.~(\ref{boxComplex}) is replaced by
\begin{eqnarray}
  \Box{U}^{A}
  +2\mu(A)\Omega^2\partial_{\varphi}{V}^{A}
  +\mu(A)^2\Omega^2{U}^{A}&=&\mbox{Real part 
    of }\left({\cal Q}^A\right)\label{BoxU}  \\
  \Box{V}^{A}
  -2\mu(A)\Omega^2\partial_{\varphi}{U}^{A}
  +\mu(A)^2\Omega^2{V}^{A)}&=&\mbox{Imaginary part of }
\left({\cal Q}^A\right)\,.\label{BoxV} 
\end{eqnarray}
The $\Box$ operator here, as in Eq.~(\ref{boxComplex}), is
$\eta^{\alpha\beta}\partial_\alpha\partial_\beta$, and the helically
symmetric time derivatives are implemented through the replacement
$\partial_t\rightarrow-\Omega(x\partial_y-y\partial_x)=-\Omega\partial_\varphi
$, so that
\begin{equation}
\Box=\partial_{\tilde{x}}^2+\partial_{\tilde{y}}^2+\partial_{\tilde{z}}^2
-\Omega^2\partial_{\varphi}^2\,.
\end{equation}
The mathematical problem of finding the fields in full general
relativity then consists of solving the partial differential equations
(\ref{boxComplex}),  (\ref{BoxU}), (\ref{BoxV}).  The post-Minkowskian
and linear approximations to general relativity follow from making the
appropriate simplifications of ${\cal Q}^A$.

\begin{figure}[ht] 
\includegraphics[width=.3\textwidth]{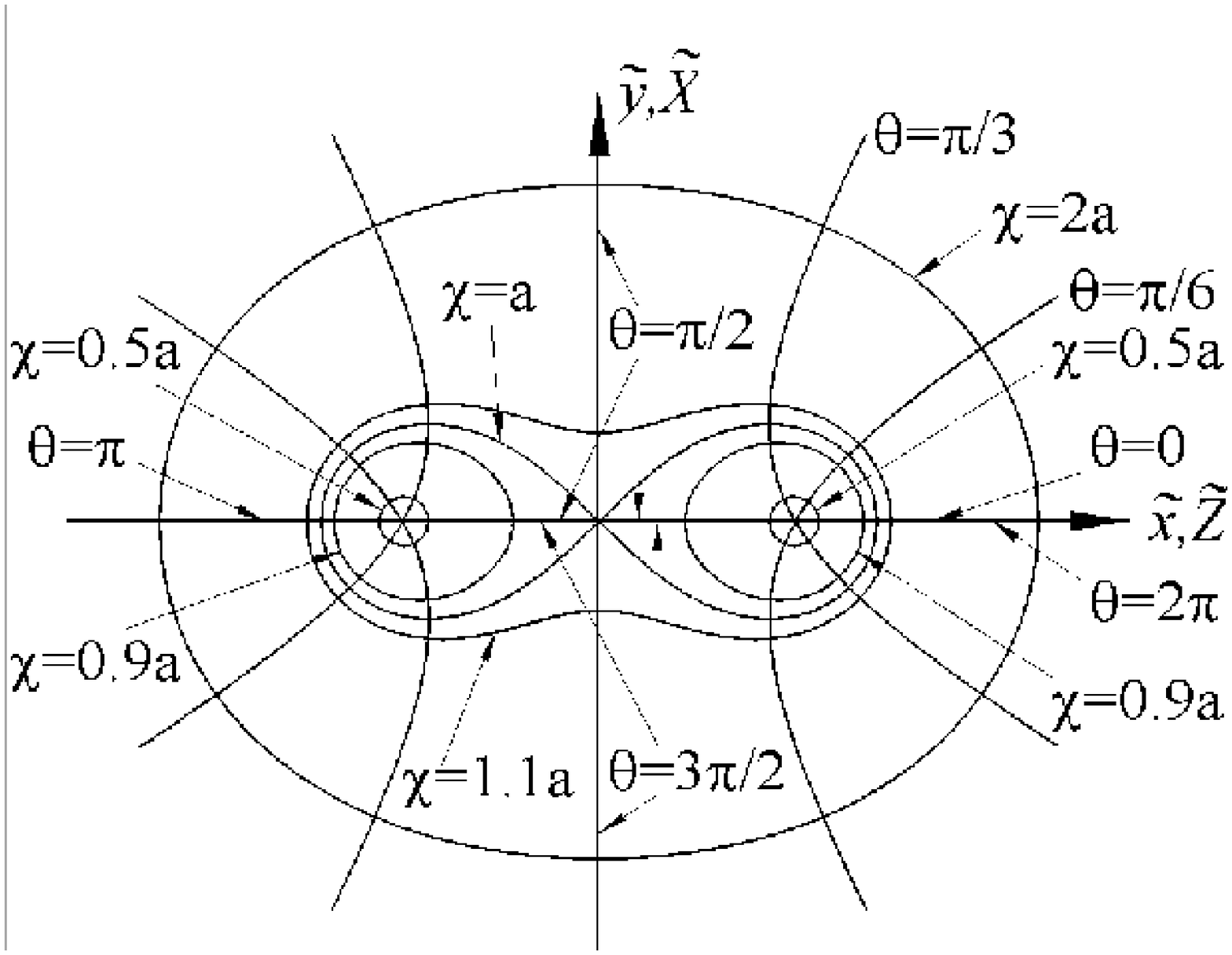}\quad\quad
\includegraphics[width=.3\textwidth]{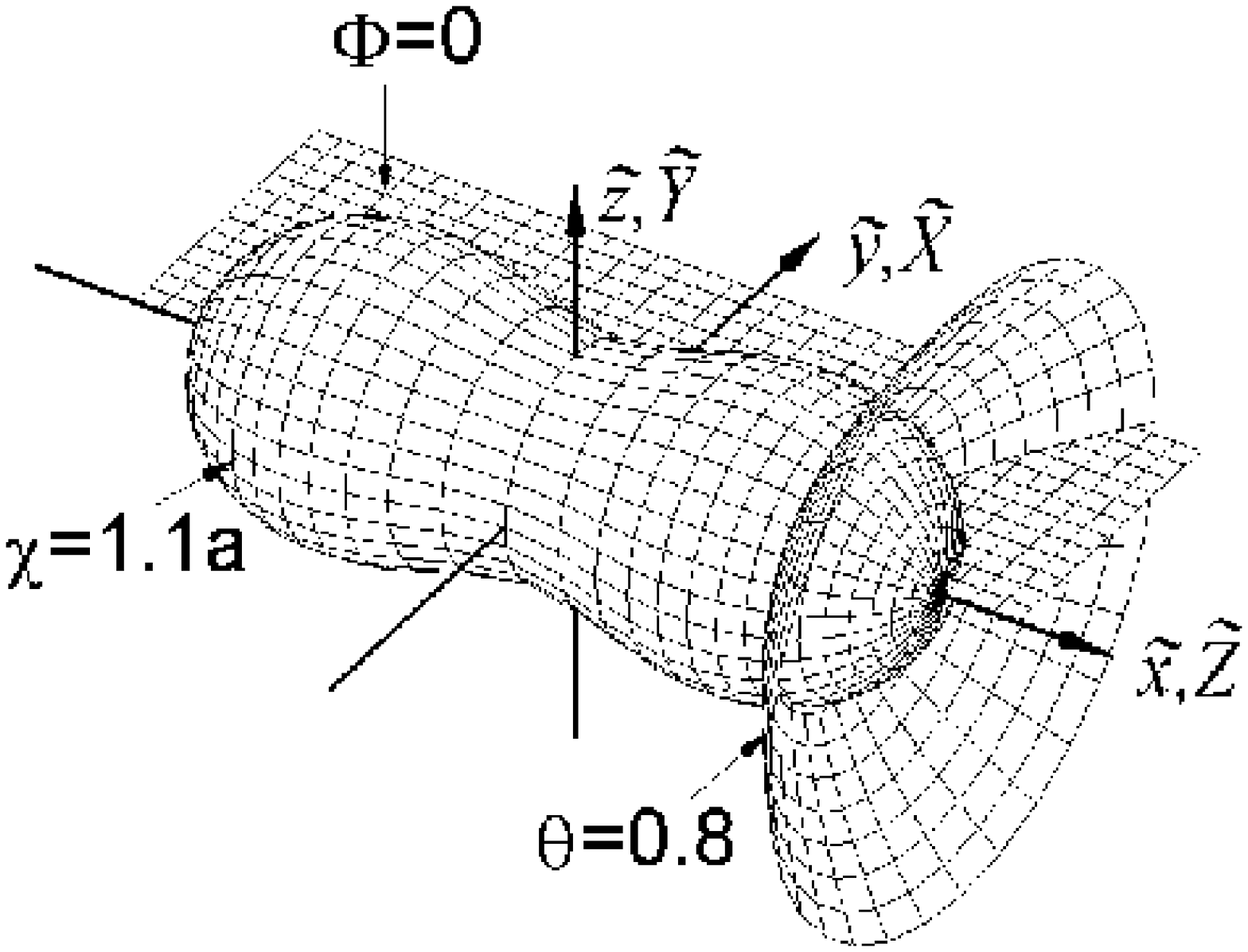} \caption{Two-center
bipolar adapted coordinates. On the left is shown curves of
coordinates $\chi$ and $\Theta $ in the $\Phi=0$ orbital plane. 
On the right are
 surfaces of constant $\chi $, $\Theta$, and
$\Phi$.  
At large $\chi/a$ the adapted coordinates $\Theta$,$\Phi$
are spherical polar angles with respect to the corotating coordinates
$\widetilde{X},\widetilde{Y},\widetilde{Z}$,
a permutation of the
coordinates
$\widetilde{x},\widetilde{y},\widetilde{z}$,
discussed in the text.
\label{fig:2dros}}
\end{figure}
Computational solutions of the field equations are carried out in an
``adapted'' corotating coordinate system $\chi$,$\Theta$,$\Phi$
illustrated in Fig.~\ref{fig:2dros}.  These coordinates are two-center
bipolar coordinates, defined in terms of the corotating coordinates
$\widetilde{x},\widetilde{y},\widetilde{z}$. The two ``centers'' are
the coordinate points $\widetilde{x}=\pm a$,
$\widetilde{y}=\widetilde{z}=0$, points that roughly represent the
locations of the black hole sources.  At large values of $\chi/a$, the
adapted coordinates approach ordinary spherical coordinates with
$\chi$ the radial coordinate. At small values of $\chi/a$, the adapted
coordinates become a reparametrization of spherical coordinates
centered on the bipolar points, with the  $\chi$ coordinate approaching
$\sqrt{2a{\cal R}\;}$ where ${\cal R}$ is the (corotating) coordinate
distance from one of the bipolar centers, and with  $2\Theta$
approaching the angle with respect to the $\widetilde{x}$ axis.  These
coordinates are discussed in detail in Ref.~\cite{eigenspec}.

Each of the helical scalar fields is expanded in the form
\begin{equation}\label{expansion} 
\Psi=
\sum_{\ell m}
\;
a^{(A)}_{\ell m}(\chi)
Y_{\ell m}(\Theta,\Phi)\,.
\end{equation}
Due to the nature of the coordinates only a few multipoles need to be 
kept. The argument for this ``multipole filtering'' 
is given  in Ref.~\cite{eigenspec}, along with numerical results 
demonstrating its validity.

\subsection{The boundary conditions}\label{sec:bcs}

The standing wave condition is a procedure rather than a constraint,
like Dirichlet, Neuman or Sommerfeld conditions. This procedure is applied in the
weak wave zone at some value $\chi_{\rm max}$
of $\chi$ that is many times larger than $a$, and much larger than 
a wavelength ($1/\Omega$).  In this weak wave zone the fields can be treated as
perturbations of flat spacetime, and outgoing/ingoing conditions on
the perturbations can be imposed.  The standing wave procedure for
full general relativity starts with the system of equations
(\ref{eq:gDef}) -- (\ref{eq:llheq}).  An approximate solution for
$\h^{\alpha\beta}$ is put into the right hand side of
Eq.~(\ref{eq:llheq}), both for the explicit appearance of the
$\h^{\alpha\beta}$ terms and  through the dependence of the
$\g^{\alpha\beta}$ terms on $\h^{\alpha\beta}$.  The right hand side
is then treated as known, and Eq.~(\ref{eq:llheq}) is considered to be
a linear equation  for 
$\h^{\alpha\beta}$. 
In Ref.~\cite{eigenspec}
it was shown that in adapted coordinates  outgoing (+)
and ingoing (-) conditions on any of the helical scalars
are
\begin{equation}\label{FDMoutbcapprox} 
\frac{1}{\chi}\frac{\partial}{\partial\chi}\left(\chi\widetilde{\Psi}\right)
=\pm\Omega\,\left(\cos\Phi\frac{\partial\widetilde{\Psi}}{\partial\Theta}
-\frac{\cos\Theta}{\sin\Theta}\,\sin\Phi
\frac{\partial\widetilde{\Psi}}{\partial\Phi}
\right) \,.
\end{equation}
These conditions are used to compute both outgoing and ingoing solutions
to the linear equations for each of the helical scalars
$\widetilde{\Psi}$. The ingoing and outgoing solutions are then
averaged, the result is taken as an improved approximation for the
field, and the process is iterated.

The convergent result, which we call the standing wave solution, is an
exact (numerical) solution of the nonlinear field equations with no
net flow of energy inward or outward. The field equations could also
be solved for a nonlinear outgoing solution, or a nonlinear ingoing
solution. In principle, for nonlinear equations, the standing wave
solution is {\it not} the average of the ingoing and outgoing
solutions. In practice, however, it is an excellent approximation for
the following reason, which we call ``effective linearity.'' The
ingoing and outgoing solution in the strong field region is negligibly
affected by the boundary conditions in the strong field zone, so that
the average of the ingoing and outgoing solution there must be the
same as the standing wave solution.  In the weak wave zone, the
solution is completely different for different radiation conditions,
but in the weak wave zone the problem is effectively linear, and again
the average and the standing wave solutions are very nearly the same.

Effective linearity means that we can treat the standing wave solution
as if it were the average, and we can decompose it into approximate
ingoing and outgoing solutions. We take the outgoing solution to be
our approximation to the solution to the physical problem.

The inner boundary is imposed on quasispherical surfaces, around the
bipolar centers, at some small value $\chi_{\rm min}$ of $\chi$. The
choice of the numerical value of $\chi_{\rm min}$ is a compromise.
The choice of a small value means that the conditions at $\chi_{\rm
  min}$ will be dominated by the presence of the black hole it
surrounds, and the other hole in the binary pair can be considered to
have a weak influence. But a small value of $\chi_{\rm min}$ means
that the computational regime $\chi_{\rm min}\leq\chi\leq\chi_{\rm
  max}$\,, includes fields strong enough to cause problems in the
convergence of the computational method.

An important point about the inner boundary conditions is made in all
previous papers on this method: the details of the inner boundary
conditions are equivalent to the details of the sources they represent
inside the $\chi=\chi_{\rm min}$ surface. But the details have almost
no effect on the fields in the wave zone, and even in the intermediate
$\chi\approx a$ zone.  We can, therefore, impose conditions at
$\chi=\chi_{\rm min}$ that represent some reasonable source, and defer
the task of adjusting the details. The wave fields, the computed
outgoing energy, etc. will be negligibly affected.

Our choice is taken to be the conditions for a single Schwarzschild
hole comoving with harmonic coordinates. This reasonable choice was
used in our post-Minkowksi computations, and hence that choice
facilitates comparisons with previous work. The actual form of the
boundary conditions have been previously reported
\cite{pmpaper} and are included here  for completeness:
\begin{equation}\label{adap-nn} 
\widetilde{\Psi}^{\bf nn}=\left(\frac{4M}{\cal R}+\frac{7M^2}{{\cal
R}^2}\right)\gamma^2
\,-\,\frac{M^2
}{{\cal R}^2} \frac{v^2\gamma^4
}{{\cal R}^2}
\frac{\chi^4}{4a^2}
\sin^2{(2\Theta)}\cos^2{\Phi}
\end{equation}
\begin{equation}\label{adap-n0} 
\widetilde{\Psi}^{\bf n0}
=\pm\,\frac{M^2}{{\cal R}^2} 
\,\frac{v\gamma^2}{{\cal R}^2}
\frac{\chi^4}{4a^2}
\sin^2{(2\Theta)}\sin{\Phi}\cos{\Phi}
\end{equation}
\begin{equation}
  \label{adap-n1R}
 U^{(n1)}=
\frac{M^2
}{{\cal R}^2}\,\frac{v\gamma^2}{{\cal R}^2
}
\frac{\chi^4}{4a^2} \sin(2\Theta)
\cos{2\Theta}\cos{\Phi}
 \end{equation}
 \begin{equation}\label{adapn-n1I}
 V^{(n1)}=\left[-\,\left(
\frac{4M}{\cal R}+\frac{7M^2}{{\cal R}^2}\right)
v\gamma^2 +
\frac{M^2
}{{\cal R}^2}\,\frac{v\gamma^4}{{\cal R}^2
}
\frac{\chi^4}{4a^2} \sin^2(2\Theta)\cos^2{\Phi}
\right]{\rm sgn}[\cos{\Theta}]
 \end{equation}
\begin{equation}\label{adap-00} 
\widetilde{\Psi}^{\bf 00}=\left(\frac{4M}{\cal R}+\frac{7M^2}{{\cal R}^2}\right)
 \frac{v^2\gamma^2}{\sqrt 3\;}
-\frac{M^2
}{\sqrt{3\;}{\cal R}^4}\,
\frac{\chi^4}{4a^2}
\left(1+(\gamma^4-1)\sin^2{2\Theta}\cos^2{2\Phi}
\right)
\end{equation}
\begin{equation} \label{adap-20}
\widetilde{\Psi}^{\bf 20}=-\,\left(\frac{4M}{\cal R}+\frac{7M^2}{{\cal R}^2}\right)
\,\frac{v^2\gamma^2}{\sqrt{6\;}}
+\frac{M^2}{{\sqrt{6\;}{\cal R}^4}}\,
\frac{\chi^4}{4a^2}
\left(\cos^2{2\Theta}+\gamma^4\sin^2{2\Theta}\cos^2{\Phi}
-2\sin^2{2\Theta}\sin^2{\Phi}\right)
\end{equation}
\begin{equation}
  \label{adap-21R}
U^{(21)}=  \frac{M^2}{{\cal R}^4}\,
\,
\frac{\chi^4}{4a^2}
\sin{2\Theta}\cos{2\Theta}\sin{\Phi}\;{\rm sgn}[\cos{\Theta}]
\end{equation}
\begin{equation}
  \label{adap-21I}
V^{(21)}=  - \frac{M^2}{{\cal R}^4}\,   
\,
\frac{\chi^4\gamma^2}{4a^2}
\sin^2{2\Theta}\sin{\Phi}\cos{\Phi}
\end{equation}
\begin{equation}
  \label{adap-22R}
  U^{(22)}=-\left(\frac{4M}{\cal R}+\frac{7M^2}{{\cal R}^2}\right)
\frac{v^2\gamma^2}{2}
-\,\frac{M^2}{2{\cal R}^4}
\frac{\chi^4}{4a^2}\left(\cos^2{2\Theta}-\gamma^4\sin^2{2\Theta}\cos^2{\Phi}
\right)\,.
\end{equation}
\begin{equation}
  \label{adap-22I}
  V^{(22)}=
\,\frac{M^2}{{\cal R}^4}
\frac{\chi^4\gamma^2}{4a^2}
\cos{2\Theta}\sin{2\Theta}\cos{\Phi}\;{\rm sgn}[\cos{\Theta}]\,.
\end{equation}
Here, ${\cal R}$, in terms of adapted coordinates, is given by
\begin{equation}
{\cal R}^2
\equiv
\frac{\chi^4
}{4a^2
}\left[1+\gamma^2v^2\sin^2{2\Theta}\cos^2{\Phi}\right]\,,
\end{equation}
where $v=a\Omega$ is the coordinate speed of the sources and
$\gamma$ is the corresponding Lorentz factor $1/\sqrt{1-v^2\;}$.
The symbol ${\cal R}$ is the coordinate distance from a source
``point'' in a frame comoving with the source; its form 
in adapted coordinates is derived in Ref.\cite{pmpaper}.
The $\pm$ in Eq.~(\ref{adap-n0}) distinguishes betweeen 
the condition (+) for the source point at $\Theta=0$ and the 
condition (-) for the source point at $\Theta=\pi$.

\subsection{Numerical implementation}
In practice one of the issues that requires considerable care in
numerically implementing the PSW method is treatment of the boundary
conditions. Consider first the outer boundary conditions. For the most
part, numerical implementation of boundary conditions
\eqref{FDMoutbcapprox} consistent with expansion \eqref{expansion} is
straightforward. However, special care must be taken with the
non-radiative parts of $\tilde{\Psi}^{(nn)}$, $\tilde{\Psi}^{(20)}$,
and $\tilde{\Psi}^{(00)}$, as both the analytical and numerical
solutions of these fields are dominated by non-radiative modes. In
Ref.~\cite{lineareigen} the general form of the analytical solution
for thes fields in linearized theory  is reported to be
\begin{multline}
  \Psi^A=-2K\sum_{\ell=0,2,\hdots}\frac{1}{2\ell+1}Y_{\ell 0}^*(\pi/2,0)
  Y_{\ell 0}(\theta,0)\frac{r_{<}^\ell}{r_{>}^\ell} \\ \bbla
  +4K\Omega\sum_{\ell=2,4,\hdots}\sum_{m=2,4,\hdots} m
  j_{\ell}(m\Omega r_{<}) Y_{\ell m}^*(\pi/2,0) Y_{\ell
    m}(\theta,0)\text{Im}\left[ h^{(1)}_{\ell}(m\Omega r_{>})
    e^{im\varphi}\right]\,.
\label{eq:LinearGravityPsinnSolutionDivided} 
\end{multline}
Here $K$ is a constant that depends on whether $A$ is $(nn),(20)$, or
$(00)$; $j_{\ell}(x)$ is the spherical Bessel function of order
$\ell$; $h_{\ell}^{(1)}$ is the spherical Hankel function of order
$\ell$; $r_{<}=\min(r,a)$, and $r_{>}=\max(r,a)$.

Expression \eqref{eq:LinearGravityPsinnSolutionDivided} shows that
series solutions for $\tilde{\Psi}^{(n,n)}$, $\tilde{\Psi}^{(2,0)}$,
and $\tilde{\Psi}^{(0,0)}$ are dominated by the $\ell=0$, $m=0$ mode,
which decays as $1/r$ for large $r$ values. Since the radiative part
also decays as $1/r$ it is necessary to separate numerically the
radiative and non-radiative parts of the solution at large
radius. This is accomplished by setting the following boundary
condition for the monopole $a_{00}$ (i.e., the $\ell=0$, $m=0$ mode
amplitude) of these fields
\begin{equation}
  \frac{\partial}{\partial \chi}
a_{00}=-\frac{a_{00}}{\chi}\,.
\end{equation}
This is just a mathematical
statement that at large distances the dominant part
of  $a_{00}$ falls off as $1/r$.

The numerical implementation of the inner boundary conditions must
also be treated with care. It turns out that some of the conditions in
Eqs.~(\ref{adap-nn}) -- (\ref{adap-22I}) are not to be considered as 
explicit values to be set at the inner boundary, but rather as regularity
conditions on certain of the fields. (It should be understood that 
not all the fields need to have data imposed to give them a scale. 
Due to the coupling of fields in the field equations some of the fields
are given scale by the coupling to other fields, not by their own 
boundary data. 
In practice we do this by replacing
Eqs.~(\ref{adap-n1R}), (\ref{adap-21R}), (\ref{adap-21I}), and (\ref{adap-22I})  
with
\begin{eqnarray}
\frac{\partial}{\partial_\chi} U^{(n1)} & =  & 0 \\
\frac{\partial}{\partial_\chi} U^{(21)} & = & 0  \\
\frac{\partial}{\partial_\chi} V^{(21)} & = & 0  \\
\frac{\partial}{\partial_\chi} V^{(22)} & = & 0\,,
\end{eqnarray}
since this choice has already been used in Ref.~\cite{lineareigen} and
\cite{pmpaper}.

\section{Numerical method}\label{sec:nummeth}

\subsection{Discrete spherical harmonics}\label{sec:eigen}
Discrete spherical harmonics were introduced first in
Ref.~\cite{lineareigen} as an integral part of the eigenspectral
method. 
In a straightforward approach to their use, spherical harmonics of 
analytic theory would
be evaluated at discrete grid points. We have found that this leads
to unacceptable failures of orthogonality, so instead we 
have used ``discrete spherical harmonics'' which are orthogonal 
to the level of machine precision.

To understand these, 
consider a two dimensional computational grid with $N \equiv
n_{\Theta} \times n_{\Phi}$ points. The discrete spherical harmonics
$Y^{(k)}_{ij}$ are $N$-dimensional vectors which are conveniently
represented by the two indices $1 \leq i \leq n_{\Theta}$ and $1 \leq j
\leq n_{\Phi}$. Let the matrix  $L_{ab,ij}$ be the 
operator equivalent to the angular Laplace operator evaluated at
$\Theta_a$ and $\Phi_b$ . This operator is therefore defined by
\begin{equation}
\left[\sin \Theta \nabla_{\text{ang}}^2
  f(\Theta,\Phi) \right]_{ab} \approx \sum_{ij}L_{ab,ij} f_{ij} 
\end{equation}
where $f(\Theta,\Phi)$ is an arbitrary function and where
$f_{ij}=f(\Theta_i,\Phi_j)$. The approximation symbol in the
relationship is due to the truncation error induced by the finite
difference representation of the derivatives on the left.

The discrete spherical harmonics $Y^{(k)}_{ij}$ are defined to be the
solutions of the generalized eigenvector problem 
\begin{equation} \sum_{ij}
L_{ab,ij}Y^{(k)}_{ij}=-\Lambda \sin \Theta_a Y^{(k)}_{ab}, 
\end{equation}
together with the normalization condition

\begin{equation} 
  \sum_{i}\sum_{j}
  Y_{ij}^{(k)} Y_{ij}^{(k')} \Delta \Theta \Delta \Phi = \delta_{k k'}.
\end{equation} 
Details of the construction of the matrix operator $L_{ab,ij}$ and of
the computation and properties of the eigenvectors can be found in
Refs/~\cite{lineareigen} and \cite{napthesis}.

\subsection{The use of symbolic manipulation}\label{sec:maple}
\begin{figure}[ht] 
\subfigure[]{\includegraphics[width=.4\textwidth]{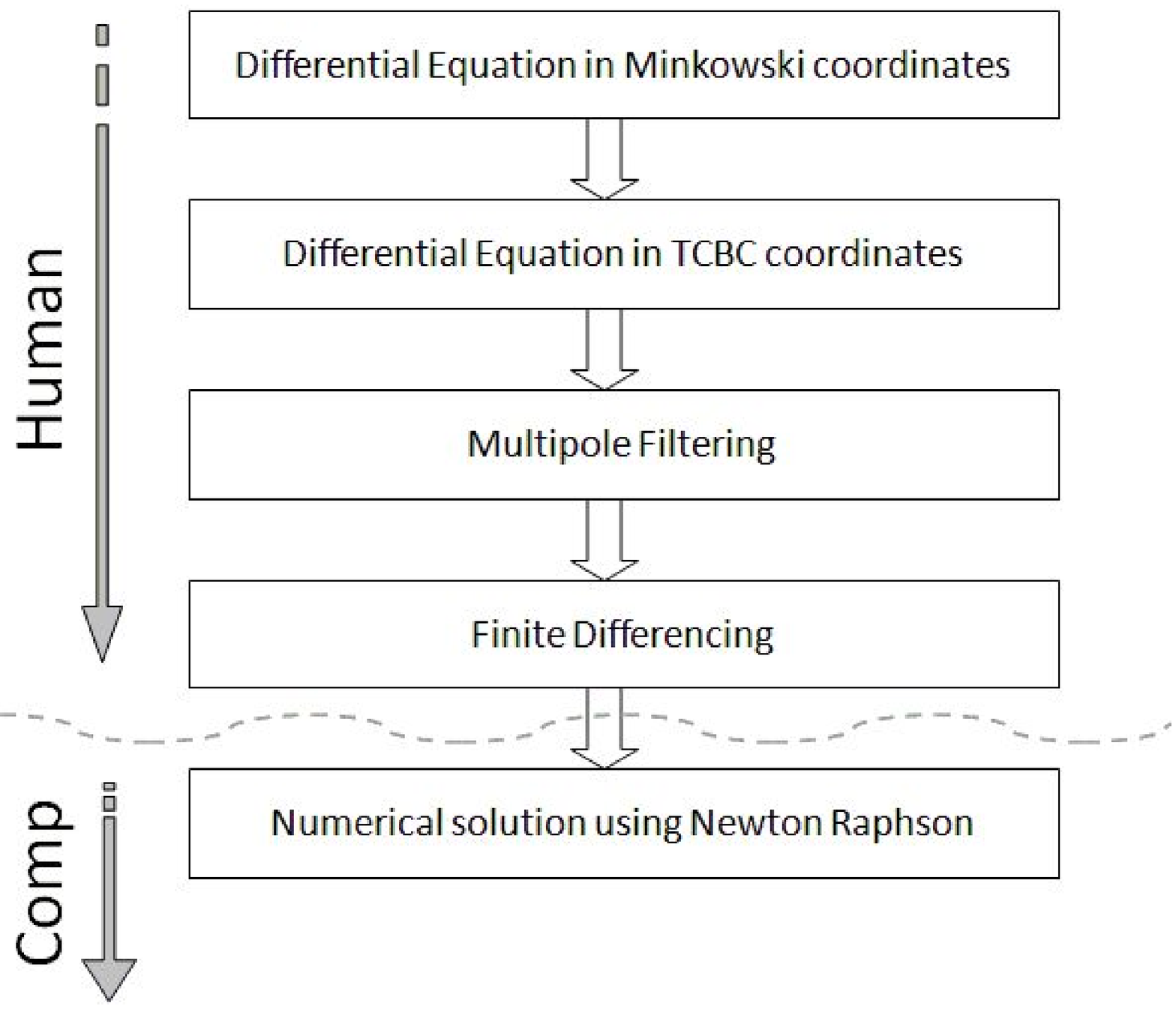}
\label{fig:ProcessFlowHuman}} \quad \quad\subfigure[]{\includegraphics[width=.4\textwidth]{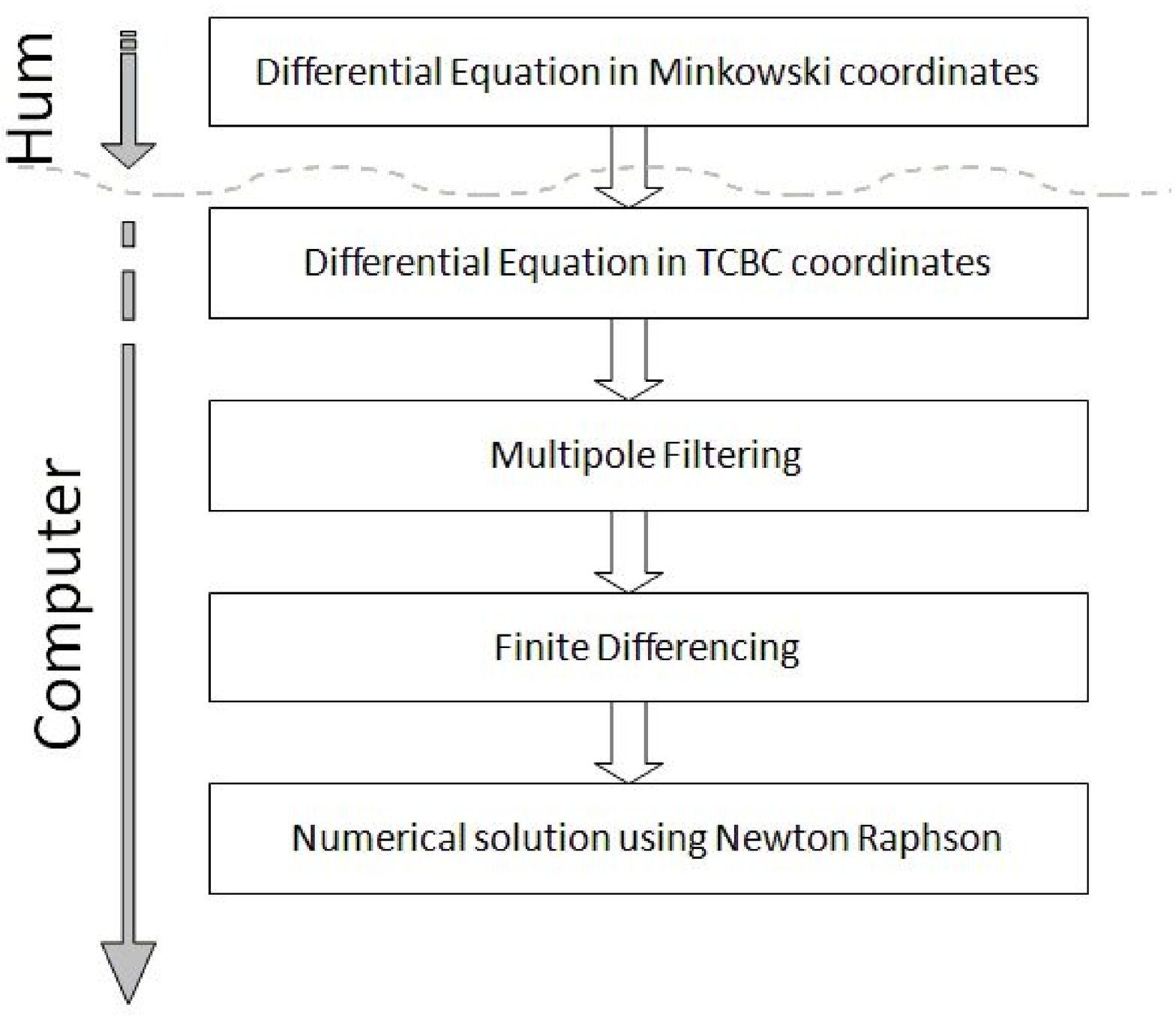}\label{fig:ProcessFlowComputer}}  
\caption{The process of implementing numerical modeling has been
  modified by the use of \emph{Maple}. The aid of a computer algebra
  system has allowed the computer to handle a great deal of the
  ``analytic'' work, significantly simplifying  the role of the
  programmer in the development of numerical code}
\label{fig:ProcessFlow}
\end{figure}

It is clear that most of the aspects involved in the computation of
fully relativistic fields with with the eigenspectral method (i.e.,
with adapted coordinates, multipole filtering, and discrete spherical
harmonics) are not conceptually difficult, but are technically
difficult to implement. The advantages of the eigenspectral method,
explained in detail in references \cite{lineareigen,pmpaper}, are
offset by an increased complexity in the form of the equations. The
several stages required by a typical algorithmic solution to a problem
within the eigenspectral method are illustrated in
Fig.~\ref{fig:ProcessFlowHuman}. During the early development of the
PSW program most of the work at these stages was carried out by a
human being, with the computer used only for the final numerical
solution of the equations.

The fact that most of the stages depicted in
Fig.~\ref{fig:ProcessFlowHuman} are algorithmic in nature suggested
the idea of developing a computational framework to deal with
the different aspects of the computation. The framework was designed
to ease the numerical implementation of any level of approximation
within the PSW model (i.e., linear, PM or fully relativistic) by
reflecting the workflow of Fig.~\ref{fig:ProcessFlowHuman}. This
framework included the implementation of several sets of tools for the
PSW project within \emph{Maple}\texttrademark, a general-purpose
computer algebra system (CAS) software package.

Three different set of tools were developed for the PSW project. The
first set corresponds to the conversion of an equation in Minkowski
coordinates $t,\tilde{x},\tilde{y},\tilde{z}$ into adapted
coordinates. The role of a second set of tools was to perform the
conversion to finite difference form of any given differential
equation. Finally a third set corresponds to tools implemented to
convert \emph{Maple} expressions into C functions. This final set
includes tools that are able to create all the different C functions
related to the projections over spherical harmonics and the
construction of the matrix system that has to be solved at each
iteration either in a perturbative or exact scheme as described in
Ref.~\cite{pmpaper}. The code generated by these tools was later
embedded into a bigger infrastructure developed in C. The C
infrastructure essentially takes care of runtime issues, such as the
allocation of memory and the interface of advanced numerical routines
in LAPACK for matrix inversion and eigenvector computation for the
routines generated by \emph{Maple}.

The implementation process for a given model within the
PSW-eigenspectral method (namely, linear gravity, post-Minkowski, or
full general relativity) is streamlined through these tools. First the
differential equations have to be provided in completely explicit
form. The first set of tools is then applied to these equations,
rendering the equations in adapted coordinates. The
equations that are output are next passed through the second set of
tools, to put them into finite difference form. Finally, the output of
this process is passed through the third set of tools in order to
obtain the C code. 

The net effect of the use of these tools ``raises the bar'' for the
implementation of numerical models with the eigenspectral method, in
that most of the process is performed by the computer, as
Fig.~\ref{fig:ProcessFlowComputer} illustrates.  We emphasize that the
use of symbolic manipulation was indispensable to the last phase of
the PSW-eigenspectral program. This leads us to suggest that a similar
eclectic use of algebraic manipulation software and numerical
programming is a powerful approach not only for our problem but for
other problems in which code must be generated for complex and lengthy
mathematical expressions for which human coding is prone to
error. However it is worth noting that neither Maple nor any other
computer algebra system is able by itself can directly generate the C
expressions required for this kind of project. Although most algebra systems features parsers to generate C code most of the
time the native parsing algorithm can not deal with some subtleties. Some of these
subtleties include type specification for intermediate variables, data type signature of functions, the appropriate translation of some mathematical functions and the creation of header files among others. For the present work it became necessary to modify the native parser of \emph{Maple} in order to deal with these subtleties. In particular the syntax definition of the C parser was enlarged in order to include some mathematical functions while the parsing algorithm was modified in order to ensure the correct signature for the functions defined during the automated process. In addition some fine tuning was performed in the parser algorithm in order to avoid incorrect or unnecessary operations in the translated code, such as some type casts or incorrect array indexing. These kind of subtleties must be addressed and tailored to the occasion by any other project that wants to use a similar approach of the mixing Computer Algebra Systems and analytic tools.   

\section{Numerical results}\label{sec:numresults}

The ultimate unknowns for which we actually
solve are the coefficients $a_{\ell m}^A(\chi)$ of the expansion of
the fields $\widetilde{\Psi}^{A}$ in Eq.~(\ref{expansion}). The
equations for these unknown $a_{\ell m}^A(\chi)$ are the field
equations (\ref{boxComplex}), along with the appropriate inner and
outer boundary conditions. In the source term ${\cal Q}^{A}$ for the
field equations, the dependence on $a_{\ell m}^A(\chi)$ occurs both in
the explicit quadratic appearance of $\widetilde{\Psi}^{A}$ on the
right in Eq.~(\ref{Qterm}), and through the very complicated
dependence of $S^{\alpha \beta \kappa \nu}_{\tau \phi \lambda \mu}$ on
$\widetilde{\Psi}^{A}$ encoded in the equations (\ref{eq:gInv}),
(\ref{eq:hdef}) and (\ref{eq:Psifirst})--(\ref{eq:Psilast}).  In our
work on the post-Minkowski approximation\cite{pmpaper} only the first
dependence was kept, since inclusion of the dependence through
$S^{\alpha \beta \kappa \nu}_{\tau \phi \lambda \mu}$ would have gone
beyond the order of the approximation.  But in the present work, for
the full equations of general relativity, we cannot ignore this deeper
nonlinearity.  In fact, it is precisely this dependence that, in our formalism,
distinguishes full general relativity from the second-order
post-Minkowski approximation. The complexity of this nonlinearity,
however, leads to a practical problem.

In our work, here and previously, with nonlinear equations, the
mathematical problem to be solved was cast into the form $L(y)=F(y)$
in which $L$ is a linear operator on the unknown function, or set of
functions $y$. A solution could be sought in two ways. A direct
iteration could be used in which an approximate solution $y_{n}$ is
substituted into $F(y)$ and the linear equation $L(y_{n+1})=F(y_n)$
for a new approximation $y_{n+1}$ is solved, with appropriate boundary
conditions. Alternatively, Newton-Raphson iteration can be used.  In
this method a linear equation is found by expanding the $y$ dependence
of $F(y)$ about $F(y_n)$, with a Jacobian playing the role of the
derivative of $F$ with respect to the fields $y$.  Newton-Raphson
iteration is expected to have better convergence properties than
direct iteration, and this was indeed found to be the case in previous
work.

In principle, then, it is a Newton-Raphson scheme that is to be sought. 
But due to the complicated dependence of 
$S^{\alpha \beta
  \kappa \nu}_{\tau \phi \lambda \mu}$ on 
$a_{\ell m}^A(\chi)$
this does not turn out to be feasible.
With \emph{Maple} generated code the evaluation of the Jacobian needed
in the Newton-Raphson method becomes unreasonable long. The main
reason for this is that the code generated by \emph{Maple} is
optimized for execution time by unrolling lengthy evaluations
explicitly in several steps using several temporary variables. As a
consequence this highly optimized code is extremely lengthy in terms
of the number of code lines. As the number of terms given to
\emph{Maple} increases so does the code that must be compiled. If the
number of terms is high enough the raw code generated with the
\emph{Maple} tools becomes so lengthy that it is impossible to
compile. This has severely limited the applicability of this automated
process for Newton-Raphson methods for a fully relativistic model. For
this reason the fully relativistic model was ultimately implemented using a
direct iteration scheme.
\begin{figure}[h!]
  \centering \includegraphics[width=.6\textwidth]{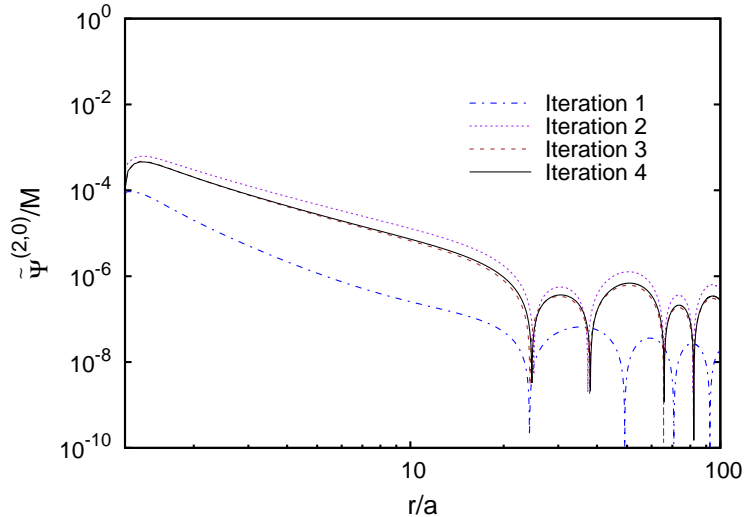}
  \caption[Typical convergence of fully relativistic
  $\tilde{\Psi}^{(2,0)}$ using a perturbative method]{Typical
    convergence of fully relativistic $\tilde{\Psi}^{(2,0)}$ using direct
    iteration. 
The field is shown as a function of coordinate radius 
along a line 
in the orbital plane approximately through the source points (i.e., 
at $\Theta=\Delta\Theta$).
The simulation parameters here are
    $\Omega=0.15$, $n_{\chi} \times n_{\Theta} \times n_{\Phi}=1750
    \times 16 \times 32$, with $\chi_{\text{min}}/a=0.50$. The
    computation includes all multipoles up to $\ell=3$ (monopole and
    quadrupole terms kept) although the monopole term $a_{00}$ is not
    included in the results displayed.}
\label{fig:FullGRConvergence}
\end{figure}

We have found that the use of direct iteration does not seem to be a
major problem. In particular, models with parameters that allowed
convergence for post-Minkowski models with Newton-Raphson iteration
have been found to converge in full general relativity with direct
iteration. One reason for this, is that we have limited our
application of the full general relativity models to those for which
the post-Minkowski approximation is valid.  As explained in
Ref.~\cite{pmpaper}, this means that $\chi_{\text{min}}$ must not be
chosen very small, to avoid very strong nonlinearities. This in turn
leads to the constraint on model parameters\cite{pmpaper}
\begin{equation}
2\sqrt{2}a \Omega \ll \chi_{\text{min}}
/a \ll \sqrt{2}\,.
\label{eq:chiMinCondition} 
\end{equation}
With this restriction on field strengths convergence has been found to
be very good.

This successful convergence is illustrated in
Fig.~\ref{fig:FullGRConvergence} for $\tilde{\Psi}^{(2,0)}$.  These
results show that convergence is attained after 3 to 4 iterations for
angular velocity $\Omega=0.15$; similar or faster convergence is found
for all smaller values of $\Omega$. We usually consider $\Omega=0.15$
to be the limit of the angular velocity for which the PSW method gives
a good approximation to the outgoing solution.

\begin{figure}[t!]
\centering
\vspace{-26pt}
\mbox{\subfigure[]{\includegraphics[width=.46\textwidth]{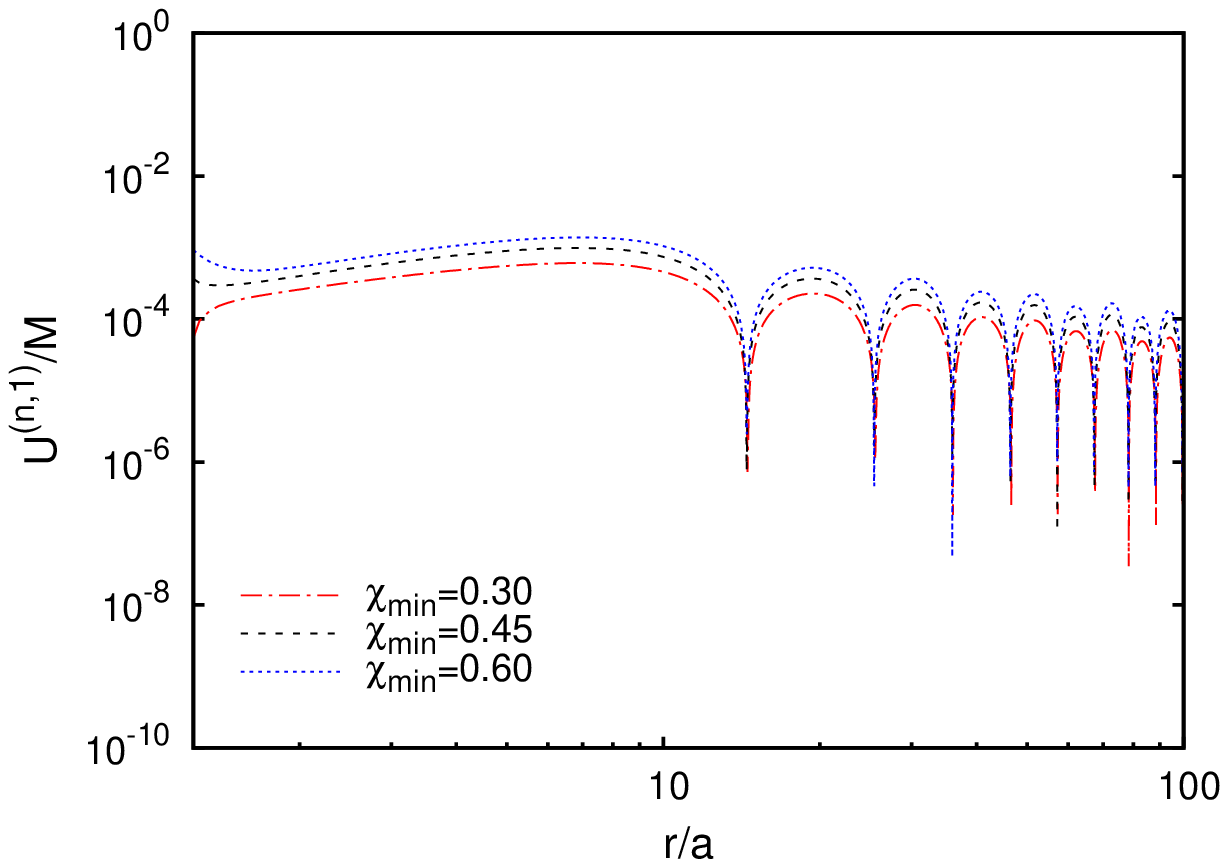} 
\label{fig:Un1FullGROmega0.15DifferentInnerChis}}}
\mbox{
\subfigure[]{\includegraphics[width=.46\textwidth]{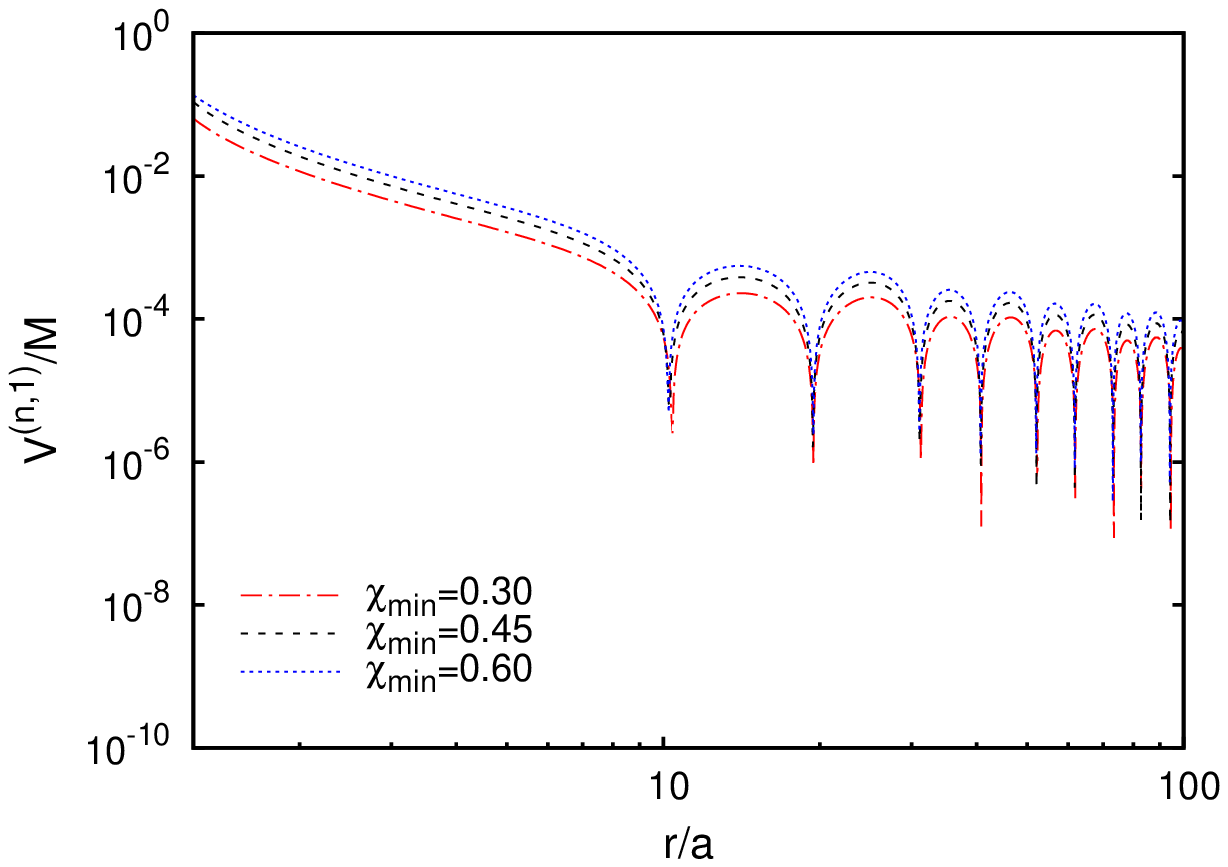} 
\label{fig:Vn1FullGROmega0.15DifferentInnerChis}}} 
\\ \vspace{-9pt}
\mbox{\subfigure[]{\includegraphics[width=.46\textwidth]{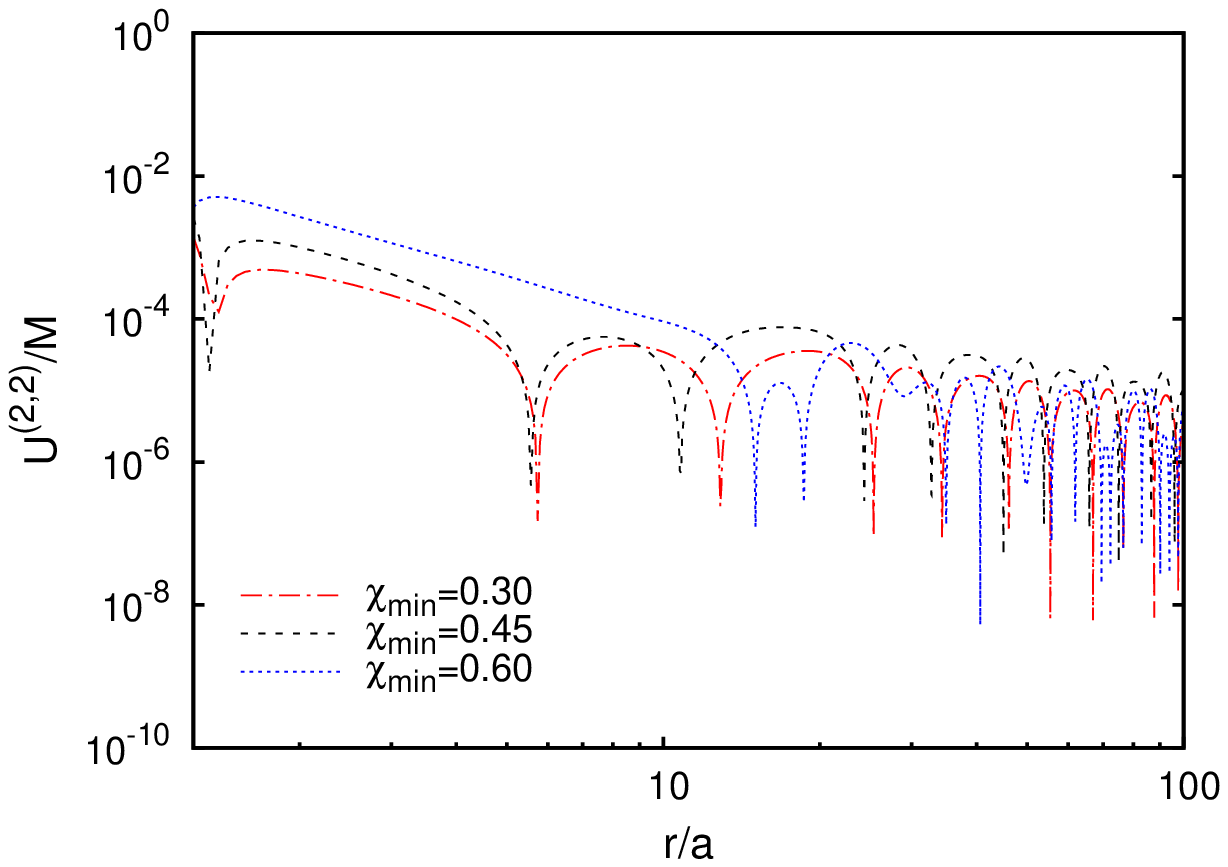} 
\label{fig:U22FullGROmega0.15DifferentInnerChis}}}
\mbox{
\subfigure[]{\includegraphics[width=.46\textwidth]{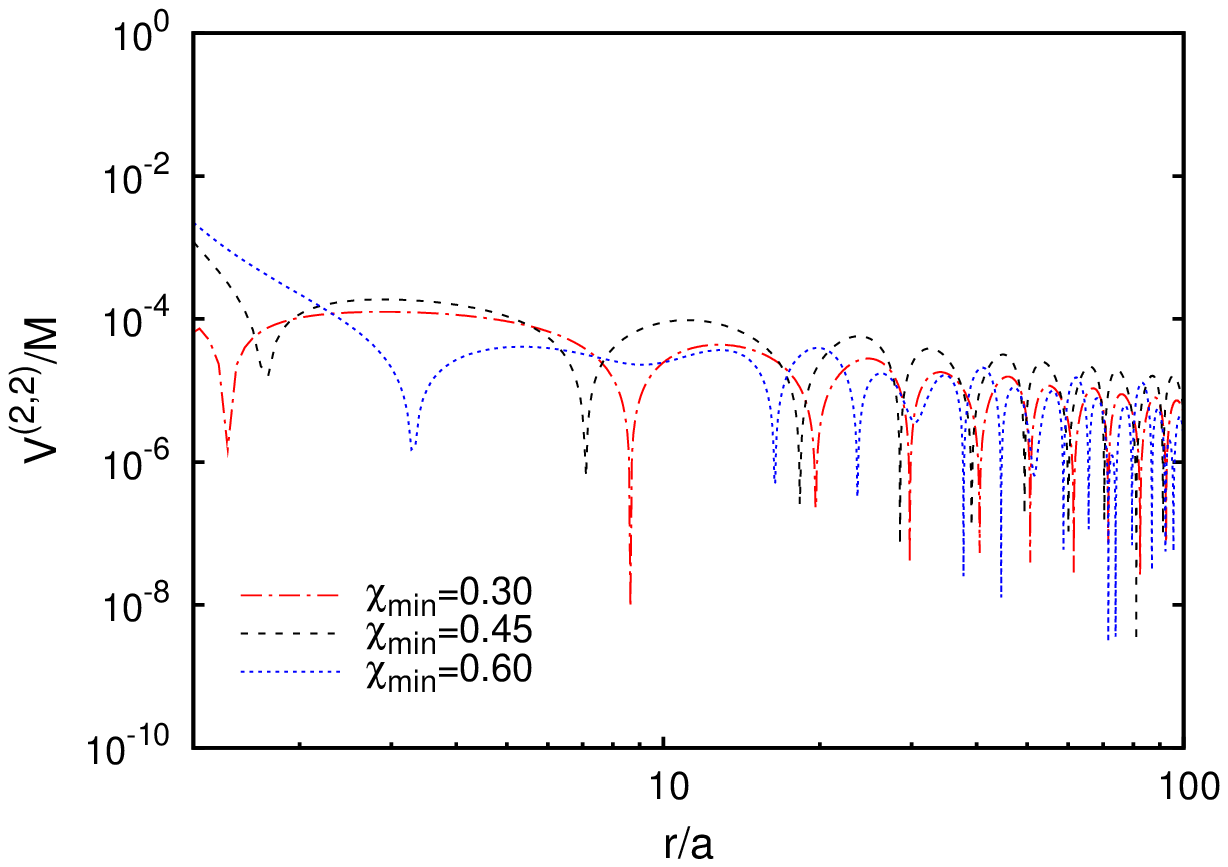} 
\label{fig:V22FullGROmega0.15DifferentInnerChis}}} 
\\ \vspace{-9pt}
\mbox{\subfigure[]{\includegraphics[width=.46\textwidth]{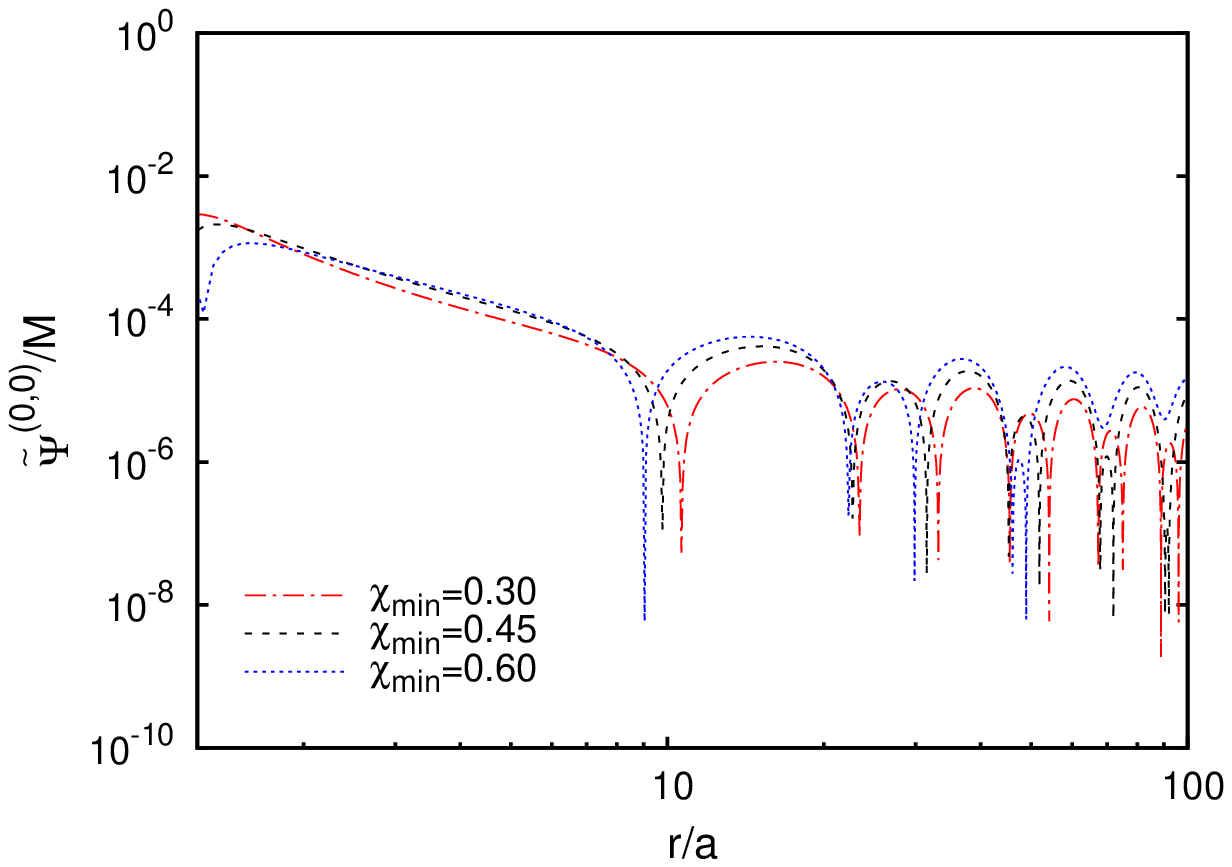} 
\label{fig:psi00FullGROmega0.15DifferentInnerChis}}}
\mbox{
\subfigure[]{\includegraphics[width=.46\textwidth]{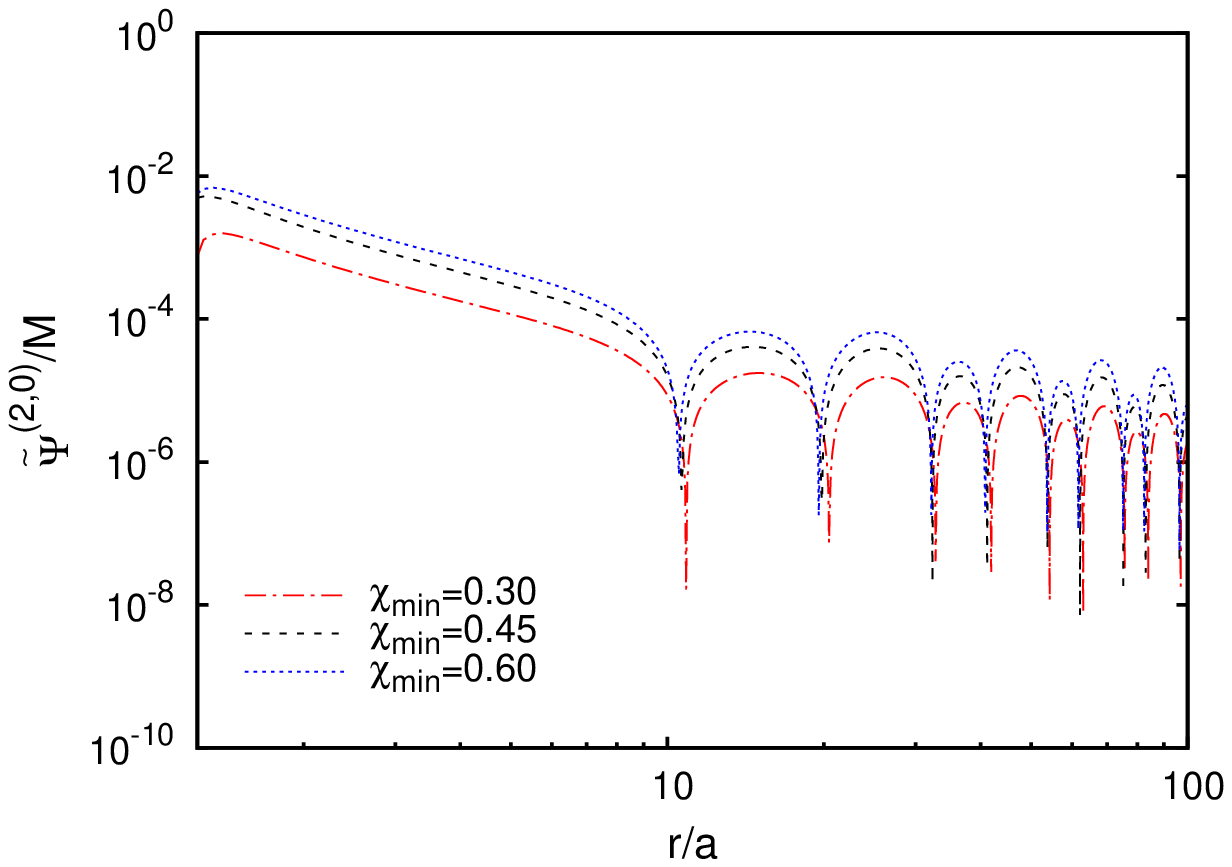} 
\label{fig:psi20FullGROmega0.15DifferentInnerChis}}} 
\\ \vspace{-9pt}
\mbox{\subfigure[]{\includegraphics[width=.46\textwidth]{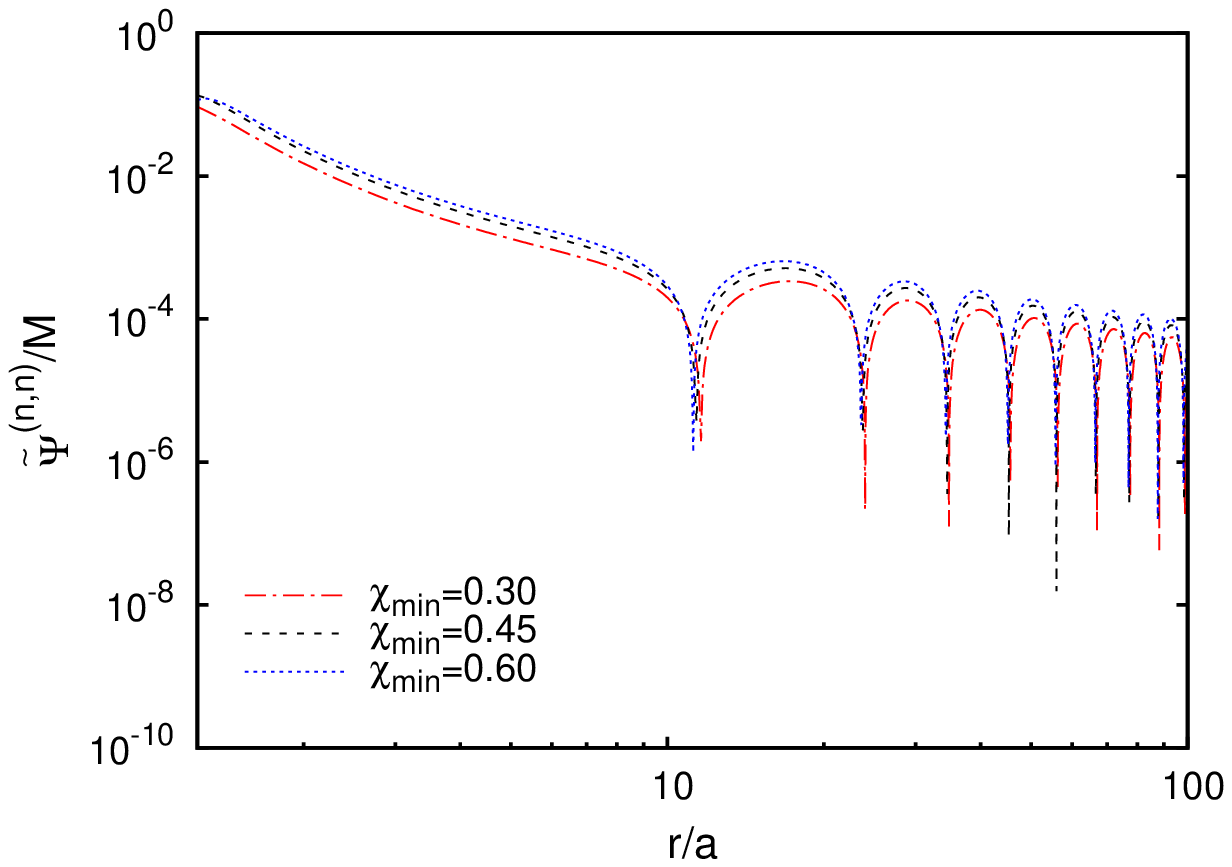} 
\label{fig:psinnFullGROmega0.15DifferentInnerChis}}}
\caption[Comparison of results for perturbative relativistic
computations as function of $\chi_{\text{min}}$ for
$\Omega=0.15$]{Comparison of results for direct iteration of relativistic
  computations as function of $\chi_{\text{min}}$ for
  $\Omega=0.15$. Fields are shown in the orbital plane along a line 
approximately through the source points.
These results reveal that the location of the inner
  boundary is less important in fully relativistic models than in
  post-Minkowski models described in \cite{pmpaper}, specially for
  $\tilde{\Psi}^{(n,n)}$. }
\label{fig:FullGROmega0.15DifferentInnerChis}
\end{figure}

A very important consideration for the success of the computations,
and of the PSW approach, is the value of $\chi_{\text{min}}$, since
this parameter determines the maximum field strength encountered in
the computation. Moreover, results that are sensitive to the choice of
$\chi_{\text{min}}$ raise questions about the underlying assumption in
the PSW approach that the details of the inner boundary condition are
not important to the fields except very near the inner boundary.
Figures \ref{fig:FullGROmega0.15DifferentInnerChis} show the results
for all fields, for a model with $\Omega=0.15$.

These results, show somewhat mixed results for the insensitivity to
the value of $\chi_{\text{min}}$. For the ``coulombic'' field
$\tilde{\Psi}^{(n,n)}$ the computed result is particularly insensitive
to $\chi_{\text{min}}$. Several of the fields show significant
diffrences in the results for the three values of
$\chi_{\text{min}}$. One would expect that the results for the largest
value of $\chi_{\text{min}}$ should have errors in representing the
boundary conditions appropriate for a ``point'' particle, since
$\chi_{\text{min}}/a=0.6$ is not really justified as a ``near source''
surface. One might expect better agreement between the two smaller
values of $\chi_{\text{min}}$, since both inner boundaries are
reasonably close to the source ``points.''  The results tend to
confirm this, though with exceptions.  Of particular interest are the
results for $U^{(2,2)}, V^{(2,2)}$, since these fields carry the
information about gravitational waves in the weak wave zone. For these
fields the results for $\chi_{\text{min}}/a=0.45$ and
$\chi_{\text{min}}/a=0.3$ agree moderately well in phase, but have
amplitude differences as large as a factor of 2. A study of the origin
of this difference will help clarify the nature of the correct
boundary conditions. It should be noted that the computations of the
fields within full general relativity have turned out to be less
sensitive to the choice of $\chi_{\text{min}}/a$ than are the
post-Minkowski computations. In some way, the inclusion of the
dependence of $S^{\alpha \beta \kappa \nu}_{\tau \phi \lambda \mu}$ on
the fields moderates the sensitivity of the results to
$\chi_{\text{min}}/a$.

\begin{figure}[t!]
\centering
\vspace{-26pt}
\mbox{\subfigure[]{\includegraphics[width=.46\textwidth]{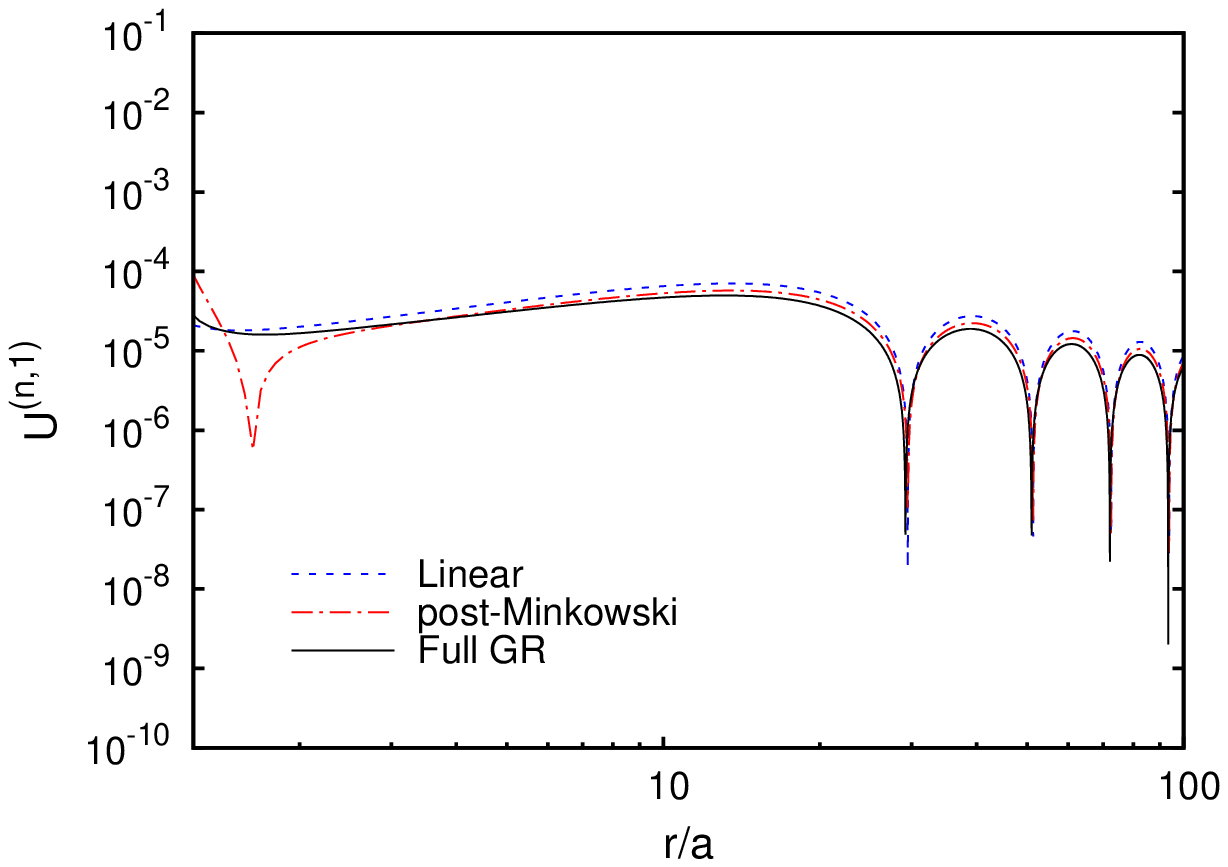} 
\label{fig:Un1LinPMNRFullGROmega0.075}}}
\mbox{
\subfigure[]{\includegraphics[width=.46\textwidth]{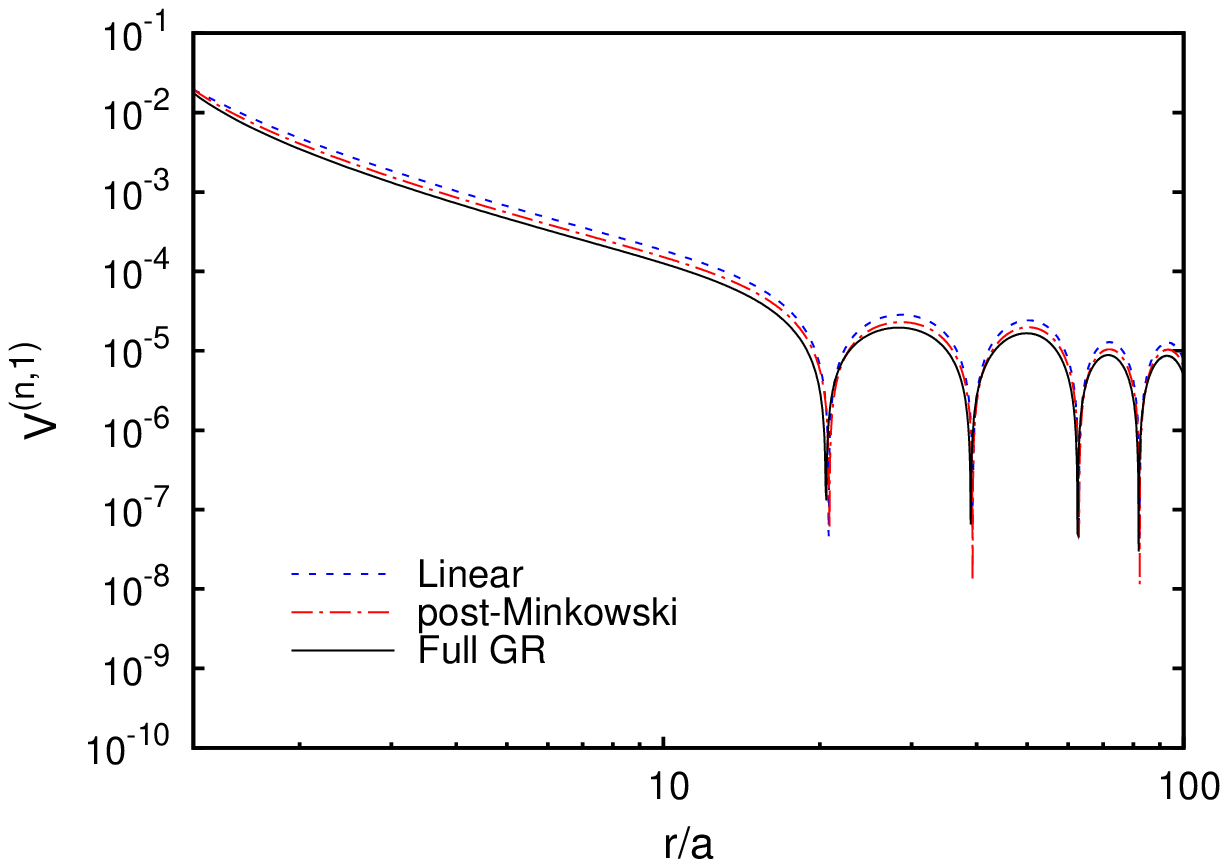} 
\label{fig:Vn1LinPMNRFullGROmega0.075}}} 
\\ \vspace{-9pt}
\mbox{\subfigure[]{\includegraphics[width=.46\textwidth]{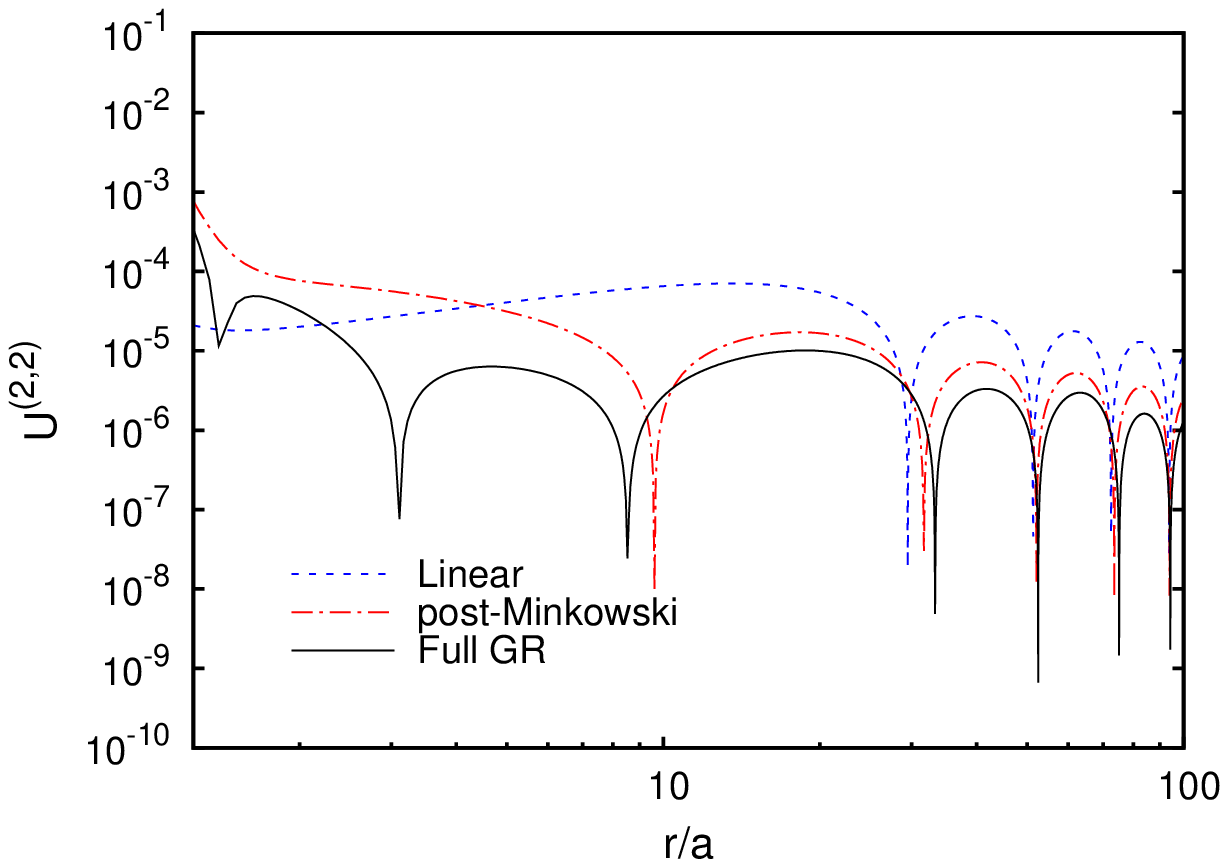} 
\label{fig:U22LinPMNRFullGROmega0.075}}}
\mbox{
\subfigure[]{\includegraphics[width=.46\textwidth]{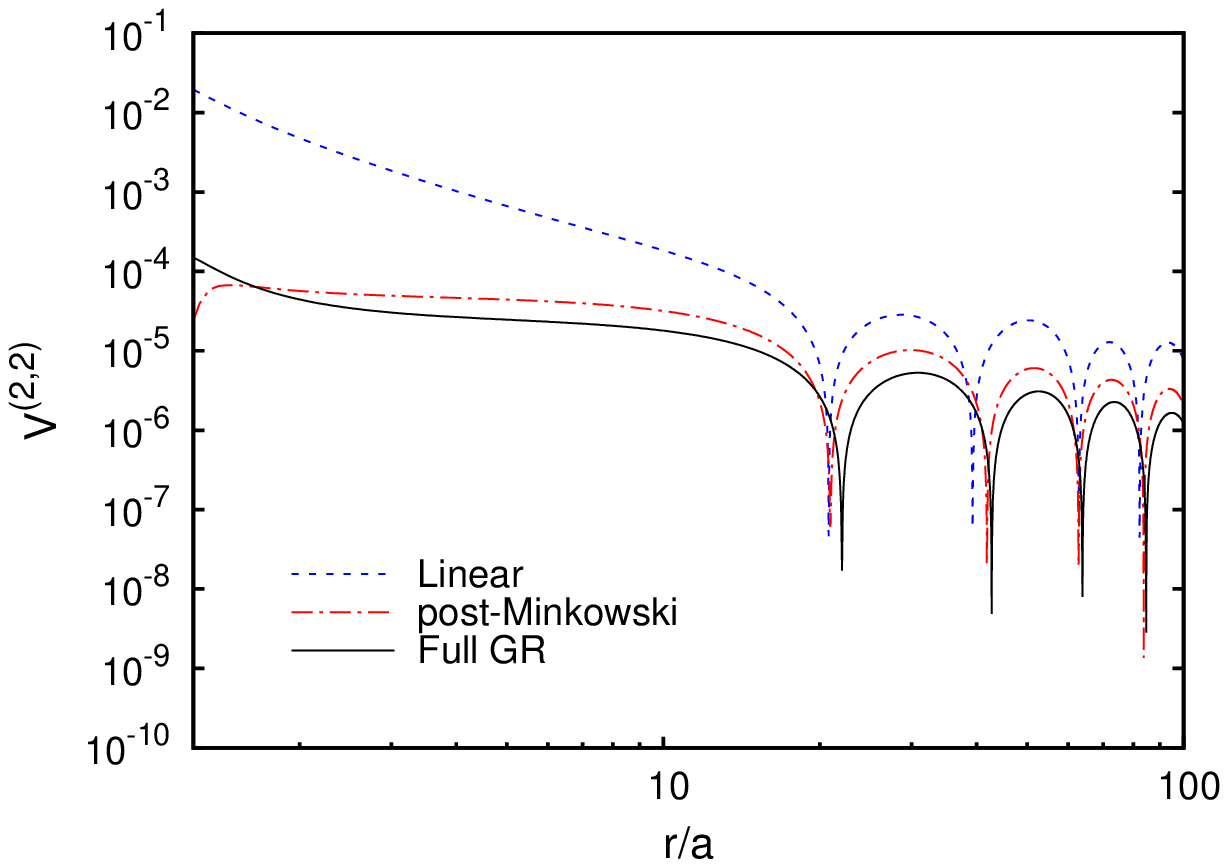} 
\label{fig:V22LinPMNRFullGROmega0.075}}} 
\\ \vspace{-9pt}
\mbox{\subfigure[]{\includegraphics[width=.46\textwidth]{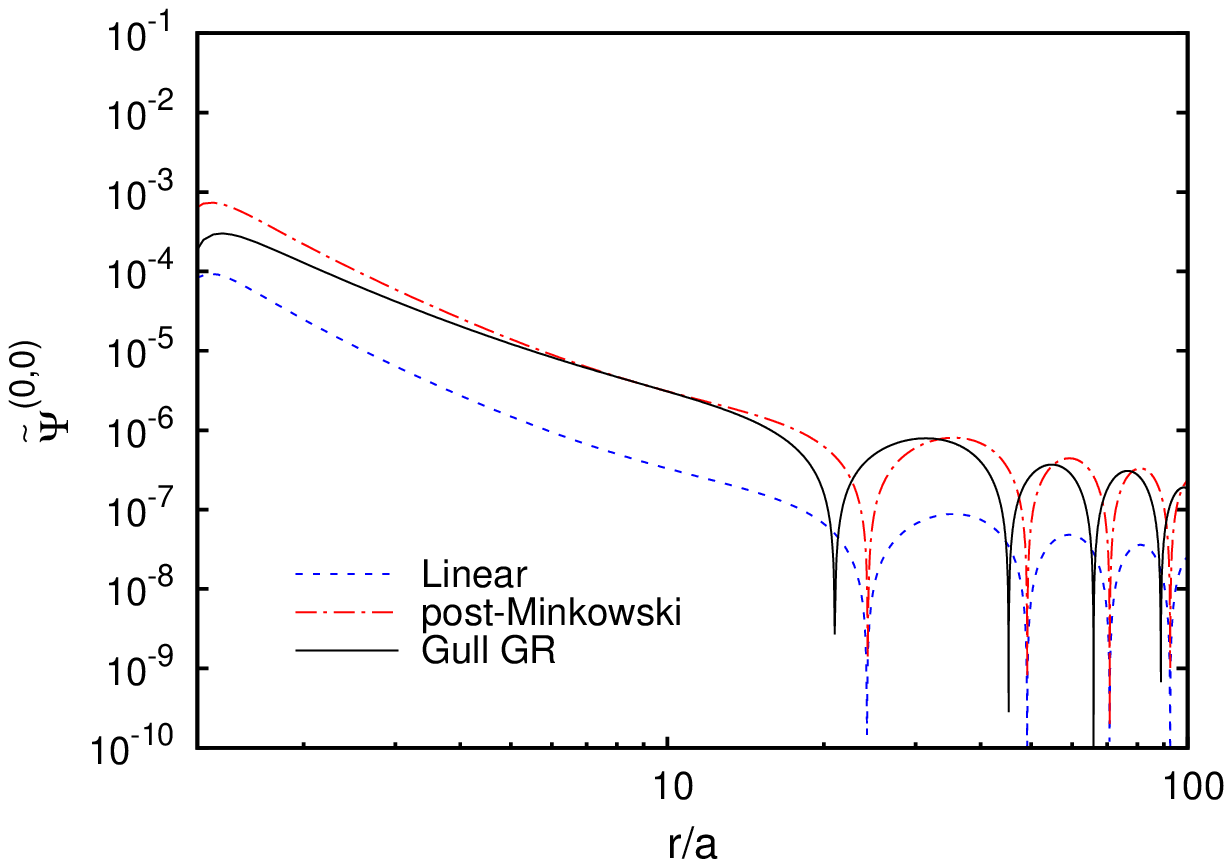} 
\label{fig:psi00LinPMNRFullGROmega0.075}}}
\mbox{
\subfigure[]{\includegraphics[width=.46\textwidth]{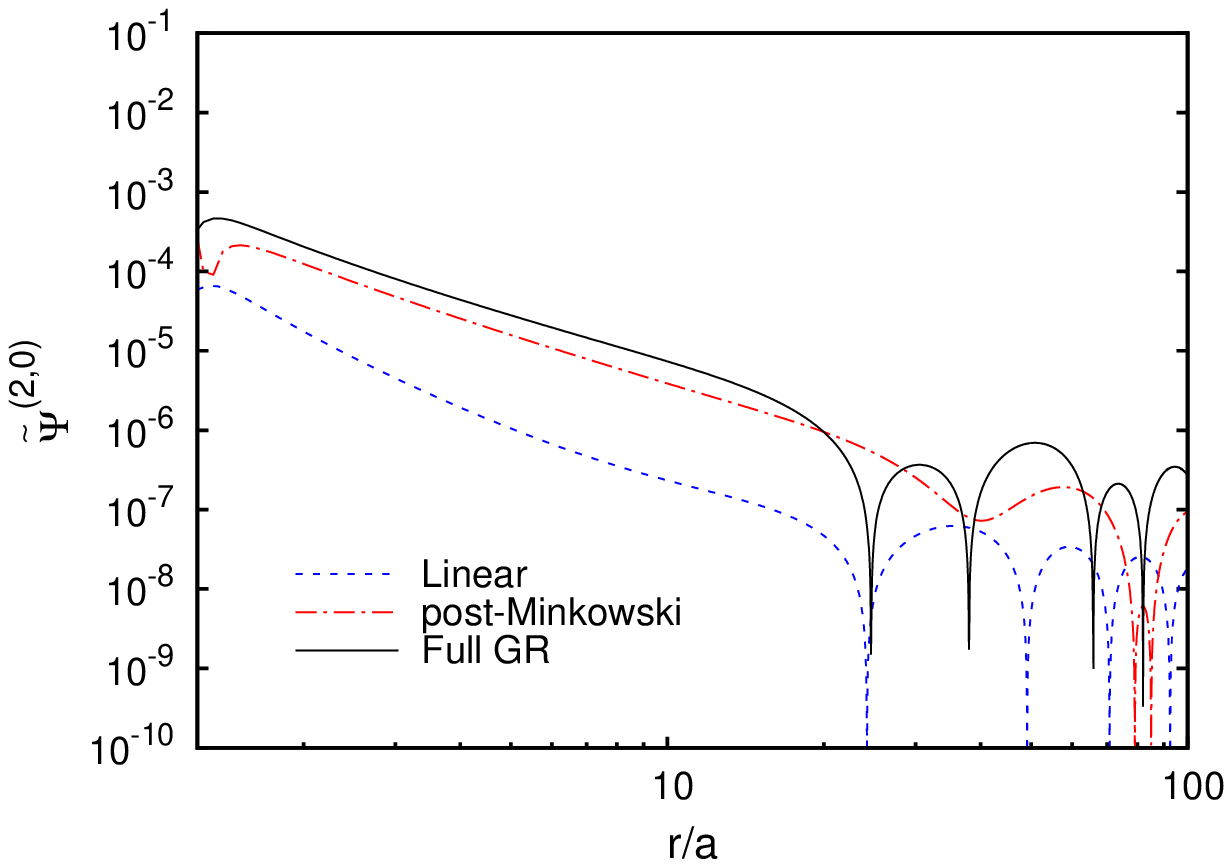} 
\label{fig:psi20LinPMNRFullGROmega0.075}}} 
\\ \vspace{-9pt}
\mbox{\subfigure[]{\includegraphics[width=.46\textwidth]{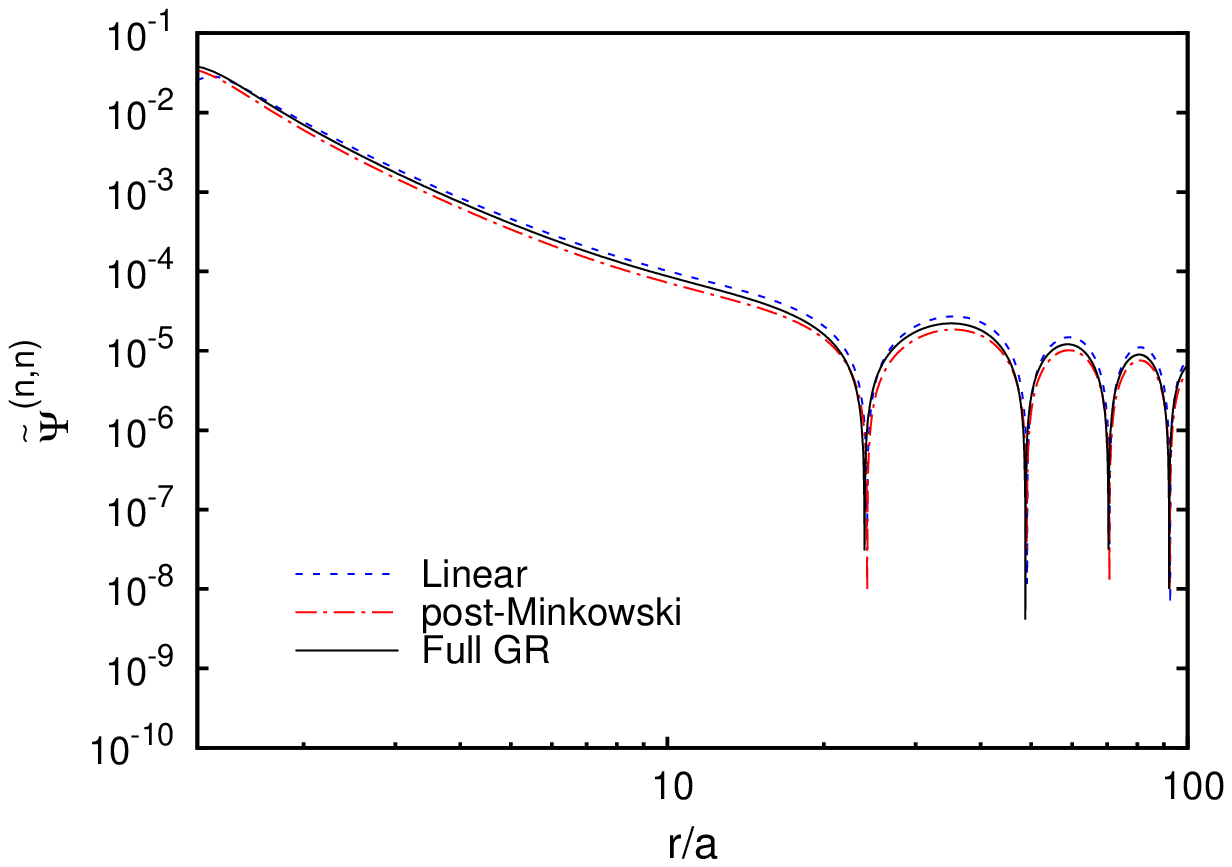} 
\label{fig:psinnLinPMNRFullGROmega0.075}}}
\caption[Comparison between linearized, post-Minkowski and fully
relativistic computations with $\Omega=0.075$]{Comparison between
  linearized, post-Minkowski and fully relativistic computations with
  $\Omega=0.075$. The computations used a numerical grid of $n_{\chi}
  \times n_{\Theta} \times n_{\Phi} = 1750 \times 16 \times 32$, with
  $\chi_{\text{min}}=0.40a$ and $\chi_{\text{max}}=100a$.  }
\label{fig:LinPMNRFullGROmega0.075}
\end{figure}

Another type of comparison is presented in
Figs.~\ref{fig:LinPMNRFullGROmega0.075}, the comparison of
computations based on linearized, post-Minkowski and fully
relativistic models for $\Omega=0.075$ and $\chi_{\text{min}}/a=0.4$.
The monopole term $a_{00}$ is not included in the results shown for
$\tilde{\Psi}^{(0,0)}$, $\tilde{\Psi}^{(2,0)}$, and
$\tilde{\Psi}^{(n,n)}$. The results show that $\tilde{\Psi}^{(n,n)}$
and $\tilde{\Psi}^{(n,1)}$ are very well approximated within the
post-Minkowski model. All other fields differ in more or less a
significant way with respect to the post-Minkowski results, showing
that the fully relativistic contributions are of significance, even
for the moderate field strengths of these computations.  In this
connection it should be noted that the field $\tilde{\Psi}^{(n,n)}$ is
very well approximated by lower order approximations. It is only this
field that was used in the source term for the iterations in
Ref.~\cite{pmpaper}. Thus, it is not the imperfections in
$\tilde{\Psi}^{(n,n)}$ that are the basis of the inaccuracies, but
rather the suppressed dependences on additional fields.

\clearpage

\clearpage

\section{Discussion and conclusions}\label{sec:conc}

The results presented in the previous section for the fully
relativistic case show the same convergence rate as was found for the
directly iterated 
post-Minkowski approximation and show similar limitations 
in field strength as the Newton-Raphson method for the post-Minkowski 
approximation. The results also reveal that the
nonlinearities present in the fully relativistic model produce
important changes in the numerical solutions for most of the
fields. For the most important field of the post-Minkowski approximation
$\tilde{\Psi}^{(n,n)}$, it is comforting to note that there is
excellent agreement of the post-Minkowski and fully relativistic results. The fact
that $\tilde{\Psi}^{(n,1)}$ is also well represented by the post-Minkowski
computation is also a good sign. These good agreements  are expected
qualitatively since the post-Minkowski approximation is correct up to
second order in the small parameter $M/\mathcal{R}$ while
$\tilde{\Psi}^{(n,n)}$ and $\tilde{\Psi}^{(n,1)}$ are of order 1 and
1.5, respectively, in the same parameter. In this sense the
relativistic results confirms the validity of the post-Minkowski
approximation.

It is important to mention not only the strengths of the method but
also its shortcomings. First, the use of adapted coordinates gives a 
solution on 
an irregular computational grid on the Minkowski background of the
problem. As a consequence the results obtained from the computations
require a difficult interpolation if they are to be compared with
other results or used as initial data for evolution codes. Second, the
method is inherently limited in its accuracy. Higher accuracy, would
ultimately require more multipoles, and an increase in the number of 
multipoles adds to the size of the equation set to be generated 
by \emph{Maple}, and to the compilation problem.

The current accuracy, however, is sufficient for us to see in the
results presented in the previous section that there is sensitivity of
the computed fields to the location of the inner boundary
$\chi_{\text{max}}$. The inner boundary conditions used for the
computations are those for a a spherically symmetric mass point, or
Schwarzschild hole, boosted so that it is instantaneously comoving
with the rotating coordinates. (For a detailed discussion see
Ref.~\cite{pmpaper}). This is, of course, an approximation that
ignores the tidal effects of the two source points on each other and
the fact that the sources are not moving in straight lines of a
Minkowksi background.  That approximation would be adequate at
sufficiently small $\Omega$, but -- the results suggest -- not at
$\Omega=0.15/a$.

Improved inner boundary conditions are an important application of the
PSW method, since this is part of the use of the method to study
radiation reaction. But the issue of the inner boundary condition may
not be separable from the issue of the gauge condition, which might
explain the sensitivity of the results to $\chi_{\text{max}}$ even for
the relatively small angular rotation rate $\Omega=0.15/a$.  The
``gauge issue'' is the fact that we have used a specific gauge, that
of Eq.~(\ref{eq:hgGauge}), but we have not enforced or checked that
gauge in the solution. If the problem is well posed then there should
be a solution which can be transformed to this gauge.  The question of
well-posedness directs attention to the inner boundary condition, and
the multipoles at the inner boundary. The coarsest parameter in those
boundary conditions is simply the mass that we ascribe to the source,
a mass that we have linked to the angular velocity $\Omega$ through
the nonrelativistic Kepler law. One would certainly expect
relativistic corrections to that relationship, so in a sense
our solution of the fully general relativistic problem has omitted
an important source of nonlinearity.

With the adapted coordinate method we have demonstrated that the PSW
computations can be brought to a level of accuracy adequate for
studying features of the strong fields and inner boundary conditions.
It is therefore capable, in principle, of giving useful insights about
radiation reaction contained in the details of the near fields. The
method can also be used in the future to provide a quasistationary
sequence of simulation parameters (i.e., mass, orbital radius and
angular velocity). Such sequence will represent more than adequately
the slow inspiral process of the binary. The main challenge for this
approach is that it is important to show that we are comparing the
``same'' system at different radii (i.e., that some physical
quantities are conserved although energy is lost during the
process). For binary neutron stars the baryon number must be
unchanged. In the case of black holes the issue is more difficult,
since the total mass of the system decreases as energy is
emitted. However the local mass of the black holes should not change;
this local mass could be calculated at each stage of the sequence
using numerical codes designed to detect horizons locally, as
described by Ashtekar \emph{et al} \cite{ashtekar2000}.

The current work has solved the problem with the accuracy
that was sought, and a broader lesson might lie in that success.  The
success was achieved through the combination of a Computer Algebra
System such as \emph{Maple} and numerical programming. This
combination can be a powerful approach to some problems that requires
not only fast and sophisticated numerical algorithms but also a fast
and sophisticated implementation process. The lesson in the PSW
program is that the use of a highly analytical approach to the problem
(adapted coordinates and multipole filtering) translated into an
increased complexity in the algebraic form of the equation. While
humans are prone to errors when manipulating lengthy expressions,
computers are not. Leaving the details of the manipulation and coding
to an algebraic manipulation language is a natural choice for this
sort of problems. Of course such an approach involves the development
of tools adapted to the needs of a particular project within
\emph{Maple} or any similar development platform, which can be quite
challenging. However, once the tools are developed and benchmarked the
correctness of the numerical implementation is guaranteed, which is a
highly desirable feature for projects involving complex numerical
implementations.

\section{Acknowledgment} 
We gratefully acknowledge the support of NSF grants 
 PHY 0554367,
and NASA grant ATP03-0001-0027.
We thank Benjamin Bromley, Christopher Beetle,  
John Friedman and Stephen Lau for many useful discussions and suggestions.


\begin{thebibliography}{10}
\bibitem{notrelated} The coordinates $X,Y,Z $ here have no relations
  to the Minkowski coordinates $X,Y,Z $ used in
  Refs.~\cite{paperI,eigenspec,lineareigen}.
\end{thebibliography}
\end{document}